# Ultrasonication-Induced Extraction of Inner Shells from Double-Wall Carbon Nanotubes Characterized via In Situ Spectroscopy after Density Gradient Ultracentrifugation

*Maksiem Erkens [a], Sofie Cambré [a,*], Emmanuel Flahaut [b], Frédéric Fossard [c], Annick Loiseau [c], and Wim Wenseleers [a,*]*

[a] Nanostructured and Organic Optical and Electronic Materials (NANOrOPT), Department of Physics, University of Antwerp, B-2610 Antwerp, Belgium

[b] CIRIMAT, Université de Toulouse, CNRS, INPT, UPS, UMR CNRS-UPS-INP N°5085, Université Toulouse 3 Paul Sabatier, Bât. CIRIMAT, 118, route de Narbonne, 31062 Toulouse cedex 9, France

[c] Laboratoire d'Etude des Microstructures, CNRS-ONERA, Université Paris-Saclay, Châtillon, France

* Correspondence should be addressed to: sofie.cambre@uantwerpen.be or wim.wenseleers@uantwerp.be

**Abstract:** Even though ultrasonication is considered to be an effective method to disperse carbon nanotubes (CNTs), its devastating effects on the nanotubes are often neglected. Here, even mild ultrasonication is found to rapidly extract the inner single-wall CNTs (SWCNTs) from the outer shells of the double-wall CNTs (DWCNTs). As-synthesized DWCNTs are gently solubilized in a surfactant solution, strictly avoiding any ultrasonication, followed by two consecutive density gradient ultracentrifugation (DGU) steps to obtain a purified colloidal solution of isolated DWCNTs. The latter is carefully selected based on *in situ* resonant Raman (RRS) and fluorescence (PL) spectroscopy, measured as a function of depth directly in the ultracentrifuge tube after DGU. These purified DWCNTs are ultrasonicated in successive time steps while intermittently probing the sample *via* RRS and PL spectroscopy. These results unravel the very fast increasing yet saturating extraction mechanism that leads to the formation of fluorescing SWCNTs. A statistical high-resolution transmission electron microscopy study confirms the drastic increase in SWCNTs after ultrasonication, and evidences that ultrasonication forms SWCNTs from both the inner and outer shells of the DWCNTs. This study demonstrates how easily ultrasonication extracts SWCNTs from individually solubilized DWCNTs, unavoidably complicating any further spectroscopic studies on DWCNTs severely.

## 1. Introduction

During the past years, carbon nanotubes (CNT) have drawn increasingly more attention because of their promising electronic, optical and mechanical properties [1–3]. This might be even more the case for double-wall carbon nanotubes (DWCNT), which consist of two coaxial single-wall carbon nanotubes (SWCNT). In such a system, the inner single-wall of the DWCNT is shielded from the environment by the outer shell [4]. This gives rise to a much higher mechanical and chemical stability of the inner tube, which can lead to interesting applications that are otherwise not feasible by SWCNTs [5–8]. Despite this increased stability, we demonstrate in this study that ultrasonication has dramatic effects on the structure of individually solubilized DWCNTs that lead to the extraction of SWCNTs from the DWCNTs.

Geometrically speaking, a DWCNT can be considered as a simple composition of two individual SWCNTs. Its optoelectronic [9,10] and vibrational [11–13] properties, however, can be strongly modulated with respect to the individual SWCNTs by weak inter-wall van der Waals interactions, depending on the sub-nanometer inter-layer distance and the twist angle between the chiral vectors of the two shells. To characterize these modified properties often the same combination of optical spectroscopic techniques as for SWCNTs is used, such as absorption, resonant Raman (RRS) and photoluminescence-excitation (PLE) spectroscopy [14,15]. Other techniques such as high-resolution transmission electron microscopy (HRTEM) and electron diffraction are especially useful as well since they enable a robust identification of the chiral structure of both the inner and outer shell of the DWCNT [16–18].

One of the actively researched properties of DWCNTs concerns the fluorescence (PL) of the inner semiconducting tube. By now, several works have been published on how macroscopic DWCNT dispersions indeed show fluorescence signals that closely match the PL of the inner tubes [4,19–21]. This is in stark contrast, however, with other work on macroscopic samples that conclude that the inner shells of DWCNTs have a drastically quenched fluorescence [22,23]. Finally, using single-tube measurements, Levshov *et al.* [24] were able to unquestionably prove that the inner tube fluoresces, but that its quantum yield is drastically quenched by up to four orders of magnitude with respect to that of freestanding SWCNTs. This quenching effect might be assigned to a possible energy or electron transfer from the inner to the outer tube, but the exact mechanism could not yet be discriminated [20,25,26].

Despite the clear proof of Levshov *et al.* [24], the reason why previous research led to such contradictory results remains unclear. A possible explanation could be hidden in the DWCNT sample preparation process. Generally, when a macroscopic DWCNT sample is spectroscopically characterized, the as-synthesized tubes are solubilized, either in an aqueous environment using a surfactant [27,28] or in an organic solvent using polymer wrapping [29]. More than often, this happens in combination with ultrasonication to individualize the tubes adequately [4,30]. Moreover, attention should be paid to the weakly abundant, yet highly PL efficient SWCNT byproducts of the DWCNT synthesis [31,32]. If these SWCNTs are not removed, their fluorescence might be wrongly interpreted as the PL of the inner tube. To this extent, it is crucial to first purify the DWCNTs in the macroscopic sample by, for instance, density gradient ultracentrifugation (DGU), a technique able to sort particles based upon their buoyant density inside a gradient medium [33]. DGU is therefore ideal to remove any SWCNTs present among the DWCNTs as they have a significantly lower density than the DWCNTs [23].

In 2014, Rohringer *et al.* [34] looked into if and how the aforementioned ultrasonication and DGU affects DWCNTs. They concluded that the shearing forces of ultrasonication, in addition to the well-known drastic shortening of CNTs [35], are able to open the DWCNT outer shell, after which the centrifugal forces exerted on the DWCNTs during ultracentrifugation in a gradient medium extract the inner tubes from their outer shells. According to these results, the sample preparation process itself can unexpectedly produce SWCNTs, which can again lead to a wrong interpretation of the fluorescence. Similar results were found in a TEM study by Miyata *et al.* [36] where they compared the abundance of SWCNTs in a catalytical chemical vapor deposition (CCVD) synthesized DWCNT sample before and after solution-phase ultrasonication. However, their results [36] did not show the same DGU requirement to fully extract the inner shells from the outer tubes, which is in contrast with the results of Rohringer *et al.* [34].

In this work, we guarantee a high-purity DWCNT parent solution through a multi-step DGU purification process in combination with *in situ* RRS and PL spectroscopy, measuring as a function of depth in the ultracentrifuge tube after DGU [37]. The resulting purified DWCNT sample shows, in line with Levshov *et al.* [24], a negligible fluorescence from small-diameter SWCNTs, while on the contrary an unexpectedly strong DWCNT outer shell PL signal is detected. By ultrasonicating these purified DWCNTs in sequential time steps we unravel the strikingly short timescale of the quickly saturating inner shell extraction by ultrasonication alone, proving hereby unambiguously that solely ultrasonication lies at the origin and no ultracentrifugation is needed. A statistical HRTEM study confirms that the SWCNTs formed during ultrasonication originate from both the inner and outer tubes of the DWCNTs.

## 2. Material and methods

### 2.1 Sample preparation

As this study focuses on the effects of ultrasonication on DWCNTs, it is crucial that during the sample preparation no ultrasonication is used as this is known to severely damage CNTs [27,28,38]. Therefore, in order to be in full control of the sample preparation, only pristine DWCNT powder (batch R281010), synthesized by E. Flahaut and co-workers [31], taken straight from the CCVD-reactor was used. The still present MgO-supported catalytic particles were removed by a mild hydrochloric acid treatment using the following neutralization reaction: $MgO + 2HCl \rightarrow H_2O + MgCl_2$. Since the MgO particles are much heavier than the CNTs, the mass of the DWCNT powder is approximated to be equal to the mass of the MgO particles. First, two suspensions of 100 mg of DWCNT powder in 40 ml of distilled $H_2O$ each were made in centrifuge tubes (50 ml, polypropylene, Falcon, VWR). Next, a double excess of HCl (987 µl of 37 wt./v% HCl, AnalaR NORMAPUR, 20252.290, VWR Chemicals) was added to the suspension and then mixed gently by hand, without applying ultrasonication. After one week, the water was refreshed until the pH of the suspension reached that of the distilled $H_2O$ used (here pH 5.4). This was done in multiple steps where each time the suspensions were centrifuged at 4025*g* for 15 minutes at 22 °C (Sigma 2-16KCH centrifuge, swing-out rotor), the supernatant was removed and 40 ml of fresh distilled $H_2O$ was added. In between each step, the suspensions were given one day to exchange and their pH was measured (pH probe meter, Lab 850, SCHOTT Instruments). Once neutralized, the suspensions were filtered over a 5 µm polycarbonate membrane (7060-2513, Whatman) in a Büchner setup. The filtered wet powder was rinsed with and solubilized in $D_2O$ (99.8%, Cortecnet) without in between drying, to avoid aggregation of the tubes, with 2 wt./V% of sodium deoxycholate (DOC, 99%, 218590250, Acros Organics) at a concentration of 4 mg CNT powder per ml surfactant solution, using only gentle magnetic stirring, avoiding ultrasonication [27,28]. After three months of stirring, the sample was centrifuged at 22 °C and an acceleration of 16125 *g* for 4 hours (Sigma 2-16KCH centrifuge, swing-out rotor) to sediment any undissolved species, after which the supernatant was carefully collected. This colloidal solution of DWCNTs is referred to as the parent DWCNT sample.

**2.2 Density gradient ultracentrifugation and *in situ* characterization**

To purify the solubilized DWCNTs from any SWCNTs remaining from the synthesis [31], the DWCNTs were ultracentrifuged in a gradient medium and subsequently probed using RRS and PL as a function of depth directly in the centrifuge tube, analogously to the procedure presented in reference [37]. The density gradient medium used is iohexol (5-(N-2,3-dihydroxypropylacetamido)-2,4,6-triiodo-N,N'-bis(2,3-dihydroxypropyl)isophthalamide, tradename "NycoDenz"), obtained from Axis-Shield in powder form, and was dissolved in $D_2O$ at appropriate concentrations to obtain the required density ranges. The gradients were prepared in two density phases in ultra-clear ultracentrifuge tubes (0.8 ml, 103353, Beckman). A low-density layer ($\rho$ = 1.197 g/mL, 380 µl) was superimposed on top of a high-density layer ($\rho$ = 1.274 g/mL, 355 µl). To both phases DWCNTs were added and both had a constant surfactant concentration equal to that of the DWCNT solution (2 wt./V% DOC). By tilting the ultracentrifuge tube and rotating it around its long axis to mix the two layers, a quasi-linear density gradient was achieved. In total four ultracentrifuge tubes were centrifuged (Optima Max-XP tabletop ultracentrifuge, MLS-50 swing-out rotor) during each run at 200100 *g* and 20 °C for at least 22 hours to reach isopycnic equilibrium. Like in reference [37], one of the tubes is used to relate the depth in the tube to the density of the gradient medium *via* absorption spectroscopy, see Figure S1 in the Supporting Information (SI). This relation is identical for the other ultracentrifuge tubes as shown in Figure S2 in the SI. These other three ultracentrifuge tubes were thoroughly characterized as a function of depth using *in situ* RRS and PL spectroscopy by mounting the ultracentrifuge tubes on an automated translation stage (MTS50/M-Z8, Thorlabs) [37]. For these *in situ* measurements, the spectra were recorded with a 1 mm step size in height, while the spatial resolution in PL and RRS is 1 mm and 60 µm, respectively. Shortly after the *in situ* optical characterization, to limit diffusion [37], the samples in the ultracentrifuge tubes were manually collected in fractions using a syringe based on these *in situ* measurements, as described in Section 3.2.

After collection of the DWCNT fraction, this fraction is DGU sorted a second time to further increase the DWCNT purity. Again, the density-to-depth relation is determined, the separation is monitored with *in situ* RRS and PL spectroscopy and the DWCNT fraction is collected. The gradient medium was removed from this DWCNT fraction by ultrafiltration with a 30 kDa dialysis cassette to allow for a better spectroscopic characterization further on, *i.e.* a NycoDenz Raman mode strongly overlaps with the RBMs of the DWCNT outer shells as well as the large-diameter SWCNTs. This DGU sorted DWCNT sample is referred to as the purified DWCNT sample.

**2.3 Spectroscopic techniques**

Absorption spectra were taken from 200 to 2500 nm using a Cary 5E UV-VIS-IR spectrometer. To contain the sample, 60 µl quartz microcells with path lengths of 3 mm were used. The spectra are baseline corrected with a 2 wt./V% DOC/$D_2O$ solution. However, due to the saturated $D_2O$ absorption band this correction induces a noise band around 2000 nm. Fluorescence spectroscopy was performed using a home-built setup dedicated to measure the infrared PL of CNTs. The sample is excited with a pulsed Xe-lamp (Edinburgh Instruments, custom adapted Xe900/XP920) using a 300 mm focal length monochromator (Acton SpectraPro 2355) equipped with a 1200 g/mm monochromator that has an average resolution of 6 nm in the excitation wavelength range used here. The sample's fluorescence was collected under a 90-degree angle with respect to the excitation with a 150 mm focal length spectrograph (Acton SpectraPro 2156) using a 150 g/mm grating, which has an average resolution of 16 nm, and a liquid nitrogen cooled extended InGaAs photodiode array detector with a sensitivity up to 2.2 µm (Princeton Instruments OMA V:1024/LN-2.2). The PL spectra were corrected for the emission sensitivity, including the emission filter transparency and the efficiency of the spectrograph and detector, as well as any intensity variations of the excitation pulses *via* a reference detector that measures the intensity of each pulse. Resonant Raman spectroscopy was performed with a Dilor XY800 triple spectrometer in backscattering geometry equipped with a liquid nitrogen cooled CCD detector. For the near infrared excitation wavelengths (695, 725 and 780 nm) a tunable Ti:sapphire laser (Spectraphysics, model 3900S) pumped by a 5 W $Ar^{+}$-laser (Spectraphysics, model 2020) was used. All of the Raman spectra were corrected for the setup's spectral sensitivity and the spectrometer's wavelength calibration.

**2.4 Time-dependent ultrasonication**

After two subsequent DGU separation steps, the purified DWCNT fraction was split up in two parts and stored in ultra-clear ultracentrifuge tubes to be able to ultrasonicate and characterize the extraction within the same container. One part was ultrasonicated at accumulating times using a Branson bath ultrasonicator (model 1510-MTH, 40 kHz). To verify the sonication efficiency for the exact combination of sample container and its position in the ultrasonication bath, we employed KI dosimetry to estimate the reaction rate of tri-iodide formation in water, as proposed in references [39,40]. We found a rate of $9.1 \cdot 10^{-5}$ M/s (see Figure S6 in the SI) in line with typical values reported for other bath sonicators. The other part of the sample was left unaffected for comparison. In between each ultrasonication time-step, the ultrasonicated DWCNT sample was characterized at multiple excitation wavelengths using a combination of RRS (725 and 780 nm) and PL (725, 780, 970 and 1050 nm) spectroscopy.

In order to differentiate between the different types of extracted CNT shells in this text, the following convention is adopted. An inner or outer tube of a DWCNT is referred to as the DWCNT inner or outer shell, respectively, whereas SWCNTs obtained by the ultrasonication-induced extraction of the inner or outer tube from a DWCNT are referred to as inner or outer shell SWCNTs, respectively.

**2.5 Transmission electron microscopy characterization**

The diameter distributions of the various CNT species present in the purified DWCNT sample and the extensively ultrasonicated DWCNT sample from the ultrasonication time-dependent measurements were determined using HRTEM. Here a Zeiss Libra 200 MC Transmission Electron Microscope was used that operates at 200 kV and is equipped with a monochromator and an in-column omega filter to perform high-resolution TEM with a resolution better than 2.3 Å. The slit of the monochromator was set to 3 μm to reduce the chromatic aberration of the microscope and consequently increase the contrast of the images. The in-column filter was used to remove the inelastically scattered electrons that otherwise degrade the images. The electron microscopy images are recorded with an ultrascan1000 TEM camera with an exposure time of 2 s. Since the two DWCNT samples are strongly limited in volume, have low optical densities and consist of possibly shortened nanotubes because of the ultrasonication, specific, commercially available, graphene covered lacey carbon on Cu supporting TEM grids (Single Layer Graphene TEM Support Films on Lacey Carbon, 300 Mesh Copper Grids (21710-25), PELCO®) were used. The additional graphene layer prevents the CNTs to be flushed away through the pores of the lacey carbon grid. Moreover, the graphene induces a continuous weak contrast compared to the CNTs, it keeps the CNTs in focus during acquisition and it reduces charge effects due to its metallic character. Since defects are expected to play a key role in the extraction, the DWCNTs were deposited on these grids as follows to not unwillingly damage them, see also Figure S16 in the SI. First, a droplet of the colloidal solutions was deposited onto the grids. Next, the surfactant molecules were removed from the CNTs by rinsing the grids with isopropanol in a Büchner filtration setup. The flow of isopropanol was estimated to be 30 ml over 5 minutes. After deposition, the grids were only air dried for one hour under a chemical hood. To minimize cross-contamination, the two samples were deposited with different tweezers and in between the two depositions the glassware was cleaned in a piranha etching solution ($H_2SO_4$:$H_2O_2$ (5:1)). HRTEM images were analyzed according to the procedure described in reference [16] for extracting the diameter of both SWCNTs and DWCNTs by using intensity fringe profiles along a cross-section of the tube image recorded at a focusing condition close to the Scherzer condition. Several hundreds of tubes were analyzed to get the best meaningful statistics on the diameter distribution.

## 3. Result and discussion

### 3.1 Preparation of the parent DWCNT colloidal solution

The absorbance of the parent DWCNT sample, prepared according to the careful solubilization procedure described in Section 2.1 whereby ultrasonication was strictly avoided, presents only weak CNT resonances that are superimposed on a relatively high 1/wavelength scattering background, see Figure 1. We anticipate here that the few sharp features around 1200 nm, marked by the black arrows, originate from SWCNTs present in the parent sample. Indeed, previous TEM characterization of the synthesized DWCNT sample [31] showed that both a small fraction of SWCNTs and triple-wall CNTs (TWCNTs) are synthesized together with the DWCNTs, and could thus also be present in this parent DWCNT solution. On the other hand, taking into account the conclusions of this work, it is not completely impossible that there are already contributions of extracted SWCNTs, despite the careful sample preparation.

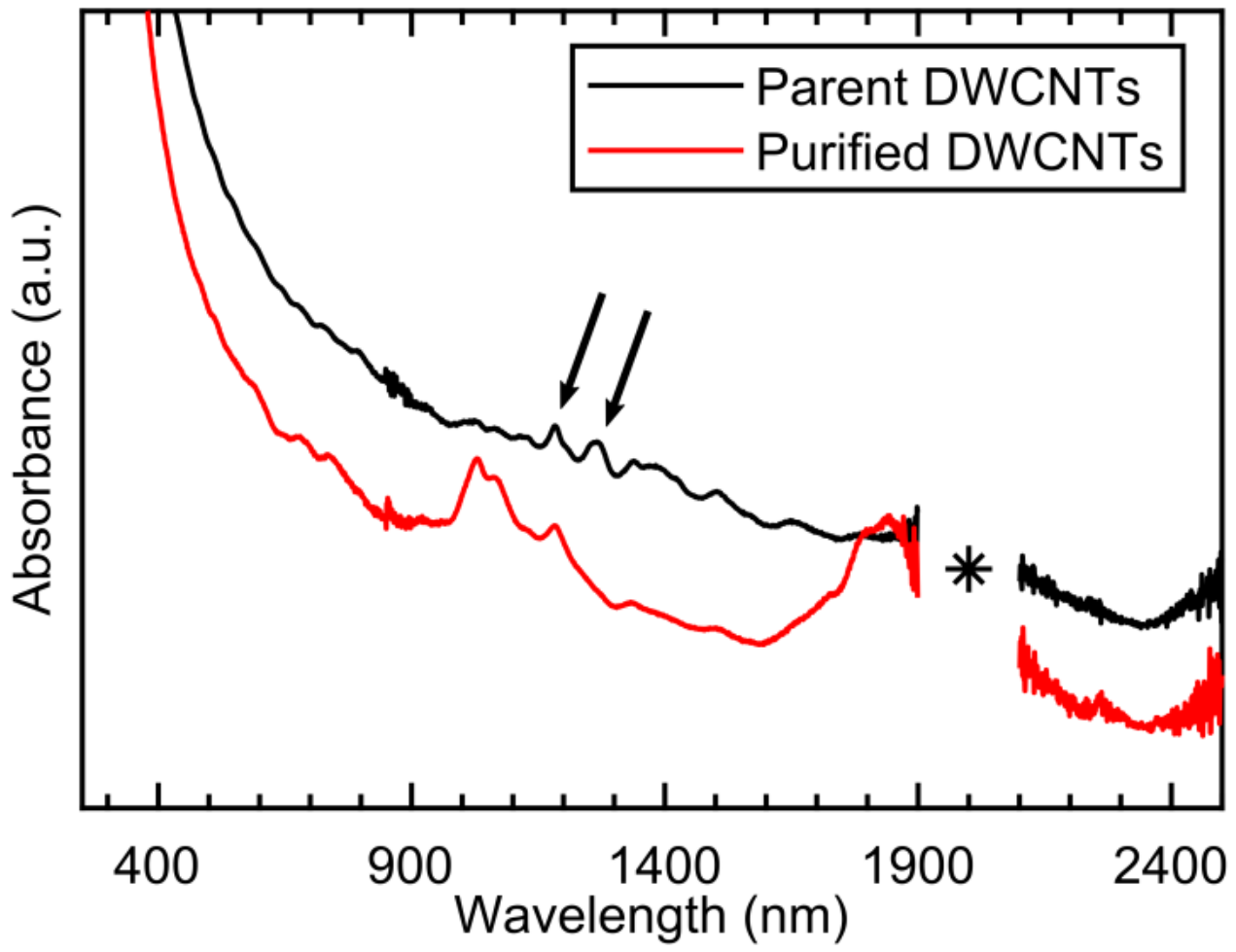

**Figure 1.** Normalized absorption spectra of the solubilized parent DWCNT sample (black trace) and the purified DWCNTs after two subsequent DGU steps (red trace). An asterisk denotes the region where the strong $D_2O$ absorption band complicates the background subtraction. The black arrows point out the possible SWCNT peaks that are present in the parent DWCNT sample. The DWCNT peaks are most pronounced in the spectrum of the purified DWCNTs.

### 3.2 DWCNT purification

To unambiguously prove that ultrasonication extracts the DWCNT inner shells, resulting in the formation of fluorescing SWCNTs, the bundles and already present SWCNTs need to be removed from the parent DWCNT sample. To this end, we apply density gradient ultracentrifugation [23,37,41]. By choosing the density gradient in the ultracentrifuge tube conscientiously, as described in Section 2.1 and in Section 1 of the SI, it is possible to physically separate the DWCNTs in the ultracentrifuge tube from the SWCNTs and bundles. After ultracentrifugation, we measure *in situ* RRS and PL as a function of depth in the ultracentrifuge tube by placing the ultracentrifuge tube on an automated translation stage in the sample compartment. This technique allows us to exactly know how the various CNT species are distributed based on their characteristic spectroscopic signals, as demonstrated previously for SWCNT DGU sorting [37,42]. Unfortunately, it is not feasible to perform fully wavelength-dependent studies within the limited time frame after DGU before significant diffusion takes place. Therefore, we chose to work with a discrete set of excitation wavelengths for RRS (670, 695, 725 and 780 nm) and PL spectroscopy (650, 730, 970 and 1050 nm). These wavelengths are chosen so that the SW- as well as the inner and outer tubes of the DWCNT distributions can be probed [14,15,43]. For the Raman measurements we focus on the frequency range of the CNT radial breathing mode (RBM) as this vibrational mode, together with the excitation wavelength, allows for the direct identification of the CNT chirality [14,15].

In Figure 2, we summarize the results of these *in situ* RRS and PL measurements as a function of depth in the ultracentrifuge tube. In Figure S3 and Figure S4 in the SI, the same results and additional PL spectra are expressed as a function of density using the density-to-depth relation from Figure S1 in the SI. In the Raman color maps we can distinguish several types of signals. First, the signal present throughout the whole depth range of the ultracentrifuge tube at 170 $cm^{-1}$ is coming from the NycoDenz gradient medium. Correction for this peak is difficult as its intensity depends on the changing density and opacity as a function of depth in the ultracentrifuge tube. Next, the RBM signals at a depth of 7.5 to 16.5 mm and in between 180 and 310 $cm^{-1}$ correspond to large- and small-diameter SWCNTs, inherently synthesized in conjuncture with the DWCNTs. The chiral indices of the small-diameter SWCNTs are indicated on the color maps in Figure 2. The larger SWCNTs, with lower RBM frequencies, are harder to identify due to the coinciding NycoDenz signal and due to their mutually strongly overlapping RBM signals.

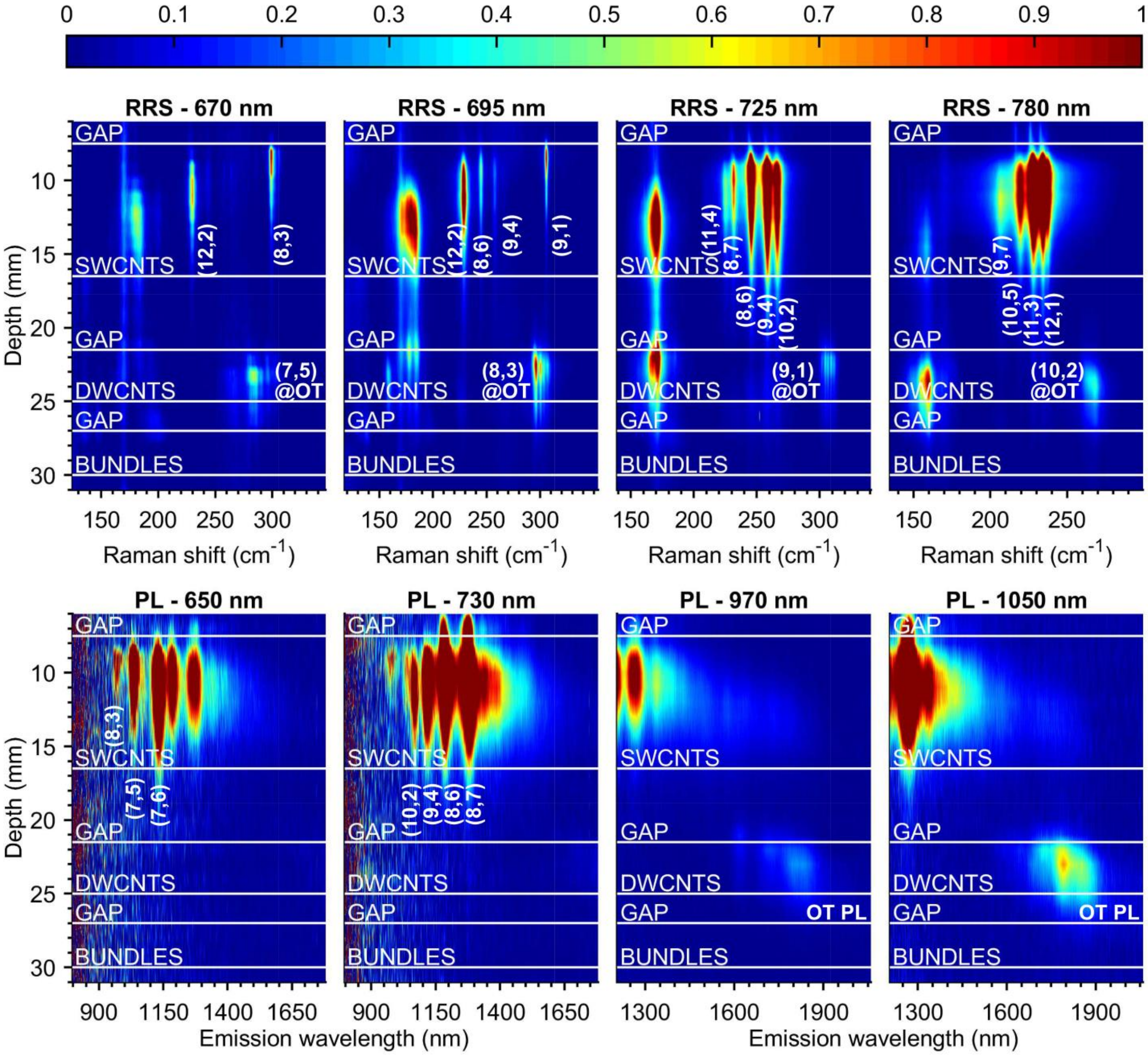

**Figure 2.** Overview of the *in situ* RRS and PL spectroscopic characterization of the DWCNT purification by DGU as a function of depth in the ultracentrifuge tube (see Figure S1 in the SI for the density-to-depth relation). The top panels show the RRS color maps measured, from left to right, at 670, 695, 725 and 780 nm. The narrow Raman signal at 170 $cm^{-1}$, present throughout the entire depth range, originates from the gradient medium. The *in situ* PL results, measured at 650, 730, 970 and 1050 nm, are shown on equal color scale but different emission axis in the bottom panels. The chiral indices of the small-diameter SWCNTs and of the DWCNT inner shells encapsulated in outer tubes (@OT) are overlaid in white. Based on the characteristic SW- and DWCNT features for all measurements, the ultracentrifuge tube is divided into fractions containing particular CNT species, indicated in white on the color maps, and collected afterwards.

We identify the DWCNTs on the color maps in Figure 2 much deeper in the ultracentrifuge tube in between 21.5 and 25 mm, which is expected as these DWCNTs have a higher buoyant density (see Figure S3 in the SI). The obtained signals around 170 $cm^{-1}$ are the RBMs of the outer tubes of the DWCNTs (and the overlapping Nycodenz-peak), while in the range from 260 to 310 $cm^{-1}$ we measure the signature splitting of an inner tube's RBM upon encapsulation in different diameter outer tubes (OT) [44]. The chiral indices of these inner tubes, identified using the Raman characterization of DWCNTs from references [43,44], are overlaid on the color maps in Figure 2. The *in situ* PL measurements confirm these SW- and DWCNT distributions. At the top of the ultracentrifuge tube, we measure the typical emission of small-diameter SWCNTs, of which the identified chiralities are overlaid on the color maps in Figure 2. Note that for excitation wavelengths 970 and 1050 nm we do not excite these SWCNTs in resonance with their $E_{22}$ transition, but within the tails of the $E_{11}$ transitions, unlike for 650 and 730 nm. Deeper in the ultracentrifuge tube, at the same depth as in the Raman color maps, a second signal is detected with emission around 1850 nm when exciting with 970 and 1050 nm. We attribute this signal to the PL of the outer tubes of the DWCNTs because of the longer wavelength infrared emission that matches with larger diameter CNTs and because of the higher buoyant density (see Figure S4 in the SI). No inner tube fluorescence is detected at this depth, as is expected from their extremely low quantum yield [24].

Based on all of the *in situ* RRS and PL measurements in Figure 2 (see also Figure S3 and Figure S4 in the SI) we assign fractions in the ultracentrifuge tubes, indicated on the *in situ* measurements in white, that correspond to the SWCNTs, DWCNTs, bundles and so-called gap fractions in between. These gap fractions are solely intended as spacer regions between the actual fractions of interest to avoid mixing of the latter during collection.

Despite leaving such a gap between the SW- and DWCNT fractions, not all the SWCNTs are successfully removed since the SWCNT signals have long tails running deeply in the ultracentrifuge tube and, hence, partially overlapping with the DWCNT fraction. Possibly, this long tail in the SWCNT density distribution could come from kinked/defective SWCNTs or from temporarily entangled SWCNTs, resulting in higher densities in comparison with individualized pristine SWCNTs [45], or from the stronger diffusion of the shorter SWCNTs causing them to overlap with the density distribution of the DWCNTs. While the first are hard to remove due to their intrinsic density, the entangled SWCNTs can be disentangled by remixing with fresh surfactant solution. Therefore, a second DGU run using an identical density gradient can help to remove the previously entangled SWCNTs and will also further reduce the diffused SWCNTs. Figure 3 shows how the DWCNT fraction of the first DGU run is selected based upon the *in situ* experiments, collected from the ultracentrifuge tube and then DGU sorted a second time. After ultracentrifugation, fractions are similarly assigned using *in situ* RRS and PL measurements at respectively excitation wavelength 725 and 1050 nm.

In comparison with the previous *in situ* DGU experiments, now the SWCNT signals in both Raman and PL spectroscopy, which are observed in the depth range of 7 to 16.5 mm, are diminished to a point where they are significantly weaker than the ones of the DWCNTs observed between 23 and 29 mm (see Figure S5 in the SI, where RRS spectra of the DWCNT fractions of DGU run 1 and 2 are compared). Hence, most of the remaining SWCNTs from the first DGU run have been successfully removed.

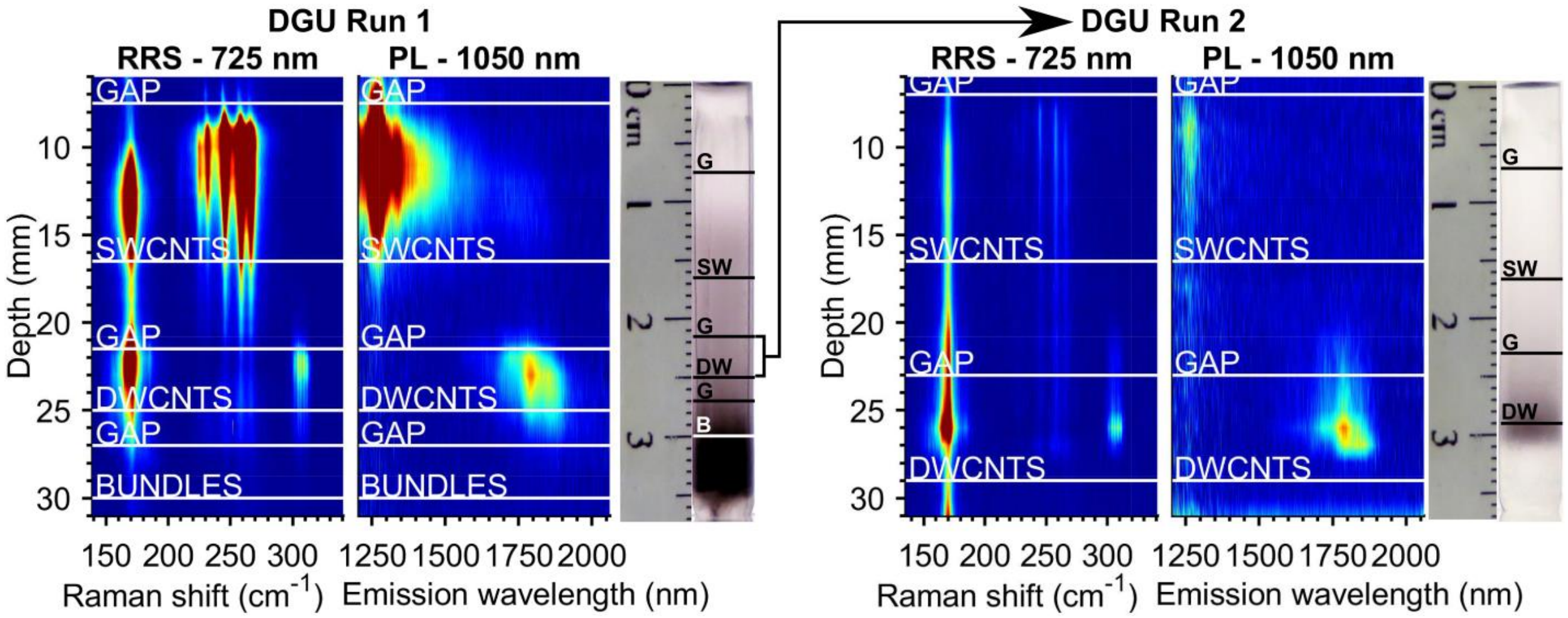

**Figure 3.** Overview of the two-step DGU purification process. After DGU sorting the parent DWCNT sample and *in situ* characterization (left panels), the DWCNT fraction is collected from the ultracentrifuge tube and then DGU sorted a second time. The *in situ* RRS and PL measurements at respectively 725 and 1050 nm (right panels) show how the DWCNT features are now significantly stronger than the SWCNT signals. Again, based upon the *in situ* characterization the DWCNT fraction is collected, which is further referred to as the purified DWCNT sample.

Nevertheless, still a small intensity from SWCNTs persists in the DWCNT density range, which are likely kinked/defective SWCNTs. Moreover, since this DGU separation was entirely based on the *in situ* spectroscopic characterization, mainly SWCNTs with a diameter up to 1.8 nm, which can be detected by RRS and PLE spectroscopy, have been removed. Any larger and therefore spectroscopically undetectable SWCNTs may still be present in the purified DWCNTs. Despite these few hard to remove kinked/defective SWCNTs or large-diameter SWCNTs a spectroscopically pure DWCNT sample is obtained after collecting the respective DWCNT fraction from the ultracentrifuge tube and dialyzing it. The spectroscopic response of this sample, which will be further referred to as the purified DWCNTs, is dominated by the spectroscopic properties of the DWCNTs. Indeed, the absorption spectrum in Figure 1 of the purified DWCNT sample (red trace) shows a remarkable increase of the ratio of the CNT peaks with respect to the background absorption, originating from the fact that the DWCNTs are well isolated, and no bundles or other impurities are present anymore due to the two-step DGU procedure. Moreover, the purified DWCNTs nicely show absorption of both the inner tube resonances (wavelength range 900 – 1300 nm, corresponding to diameters ranging from ~0.6 to 1 nm) and outer tube resonances (wavelength range 1550 – 1950 nm, i.e. diameters ranging from ~1.2 to 1.7 nm), and an absence of the previously mentioned SWCNT peaks from the parent sample (black arrows).

### 3.3 Time-dependent ultrasonication

After purifying the DWCNT sample to a level where the spectroscopic properties of the DWCNTs dominate over those of the SWCNTs, we can unambiguously demonstrate the time-dependent ultrasonication-induced inner shell extraction, as any extracted SWCNTs will lead to a clear increase in their Raman and PL signals in the purified DWCNT sample.

First, we split the purified DWCNT sample in two equal parts. One part is ultrasonicated in accumulating time steps in a bath ultrasonicator, while the other part is left unaffected as a reference. By measuring RRS and PL spectroscopy in between each ultrasonication step, we can monitor the ultrasonication-induced effects as a function of ultrasonication time. To follow these effects for both SW- as well as DWCNTs, and for different chiralities of both CNT species, we use the excitation wavelengths 725 and 780 nm for RRS, and 725, 780, 970 and 1050 nm for PL spectroscopy.

Figure 4 summarizes these ultrasonication time-dependent RRS and PL measurements (see Figure S7 in the SI for the ultrasonication time-dependent PL spectra measured at 780 and 970 nm). Before ultrasonication, the Raman and fluorescence spectra show little to no SWCNTs as is expected from the *prior* two-step DGU purification. However, these spectra change drastically as soon as ultrasonication is applied. For instance, the previously weak SWCNT RBM peaks, highlighted in red, rapidly increase in intensity, while the inner tube DWCNT Raman modes, highlighted in blue, diminish over the increasing ultrasonication time. The PL spectra show similar results as the SWCNT fluorescence (red highlight) measured at excitation wavelength 725 nm, which is completely absent before ultrasonication, quickly starts to dominate the spectrum over the DWCNT (blue highlight) outer tube's fluorescence excited at 1050 nm. Note that most of the observed effect occurs already within the first tens of seconds of the ultrasonication.

These results indicate that during ultrasonication new SWCNT signatures quickly appear in the colloidal solution, even after several seconds, while the DWCNT features disappear. Since any SWCNTs and bundles have already been thoroughly removed from the sample by the previous two-step DGU, these SWCNTs must come from the DWCNTs themselves. Our results therefore demonstrate that due to ultrasonication the DWCNT inner shell can slide out of the outer shell forming a small- and large-diameter SWCNT, respectively, in agreement with the previously suggested mechanism of Miyata *et al.* [36].

Note that while measuring the intermittent RRS and PL spectra, each time the sample is taken out of the sample compartment sonicated and returned, resulting in small alignment variations in between the successive ultrasonication steps that can be observed as an overall varying intensity of both the SW- and DWCNT signals in the spectra (see Section 2 of the SI). Therefore, to estimate the effect of the ultrasonication-induced extraction, the relative ratio between the SW- and DWCNT intensities is calculated as a function of the total ultrasonication time for RRS and PL spectroscopy separately. These relative SW- and DWCNT intensities are calculated by integrating the SW- and DWCNT signals in the RRS and PL spectra over fixed ranges for every ultrasonication time and are then averaged over all the excitation wavelengths used for RRS or PL spectroscopy, respectively. These integration ranges are indicated on the RRS and PL spectra in Figure 4 (see Figure S7 in the SI for the integration ranges of the PL measured at 780 and 970 nm) by the dashed lines and color highlights.

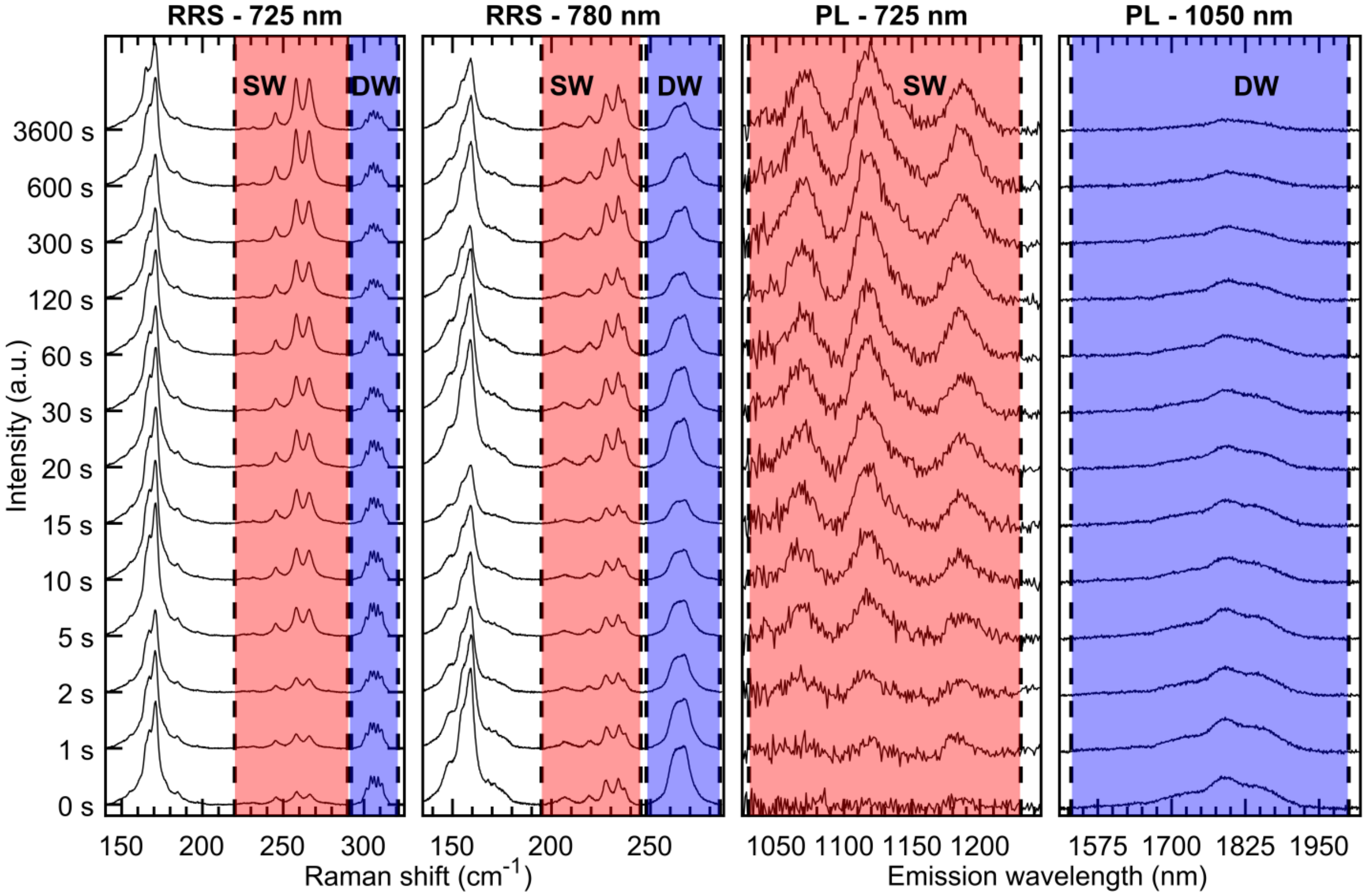

**Figure 4.** Raman and PL spectra of the purified DWCNTs measured as a function of the accumulative ultrasonication time at excitation wavelengths 725 and 780 nm for RRS, and 725 and 1050 nm for PL (other wavelengths for PL are presented in Figure S7 in the SI). The RRS and PL intensities can directly be compared between the different excitation wavelengths. Spectral regions corresponding to extracted SWCNTs and DWCNTs (inner or outer shells) are highlighted in red and blue, respectively. The ultrasonication time-dependent RRS spectra show a rapid increase in SWCNT intensity (red highlight) and a strong decrease in the DWCNT inner shell signals (blue highlight). Similarly, the PL spectra at 725 and 1050 nm show respectively a comparable increase in the SWCNT PL (red highlight) and a decrease in the DWCNT outer shell PL (blue highlight) with increasing ultrasonication time. The vertical dashed lines indicate the integration ranges used to calculate the ultrasonication time-dependent ratios between SW- (red highlight) and DWCNTs (blue highlight) as shown in Figure 5.

The resulting single-wall to double-wall intensity ratios, extracted from both PL and RRS, are displayed in Figure 5 after normalization of the data. Both ratios show a quickly increasing, yet saturating behavior, where within the first 60 s of ultrasonication already 40% of the total observed extraction is reached in PL and even 60% according to RRS. *Prior* to the normalization, the ratios, $r$, are fitted as a function of ultrasonication time, $t$, using a stretched exponentially increasing model: $r(t) = f - b\, e^{-(t/\tau)^s}$, where $f$ - $b$ and $f$ are respectively the beginning and final relative intensity ratio, $\tau$ is the extraction time constant in seconds, and $s$ represents the stretching parameter, representing the inhomogeneity of the extraction within the considered diameter distributions. The fitted time constants for both PL and RRS, shown within the legend of Figure 5, are in the order of one hundred seconds for bath ultrasonication (see Table S1 for all fitting parameters). The difference in time constants between the RRS and the PL results could be attributed to the difference between the Raman cross sections and PL quantum efficiencies of the tubes probed. In particular, a possible length-dependency of the PL quantum efficiency could also add to an apparently slower extraction according to PL. Indeed, due to ultrasonication the end-caps of the inner shells can be removed during or after extraction, or the extracted SWCNT can be cut entirely, leading to defect sites that effectively quench the SWCNT PL as the excitons can easily diffuse to these defects, which are expected because of the applied ultrasonication. Therefore, only a fraction of the extracted inner shell SWCNTs will be observed with PL, resulting in a seemingly slower extraction compared to RRS.

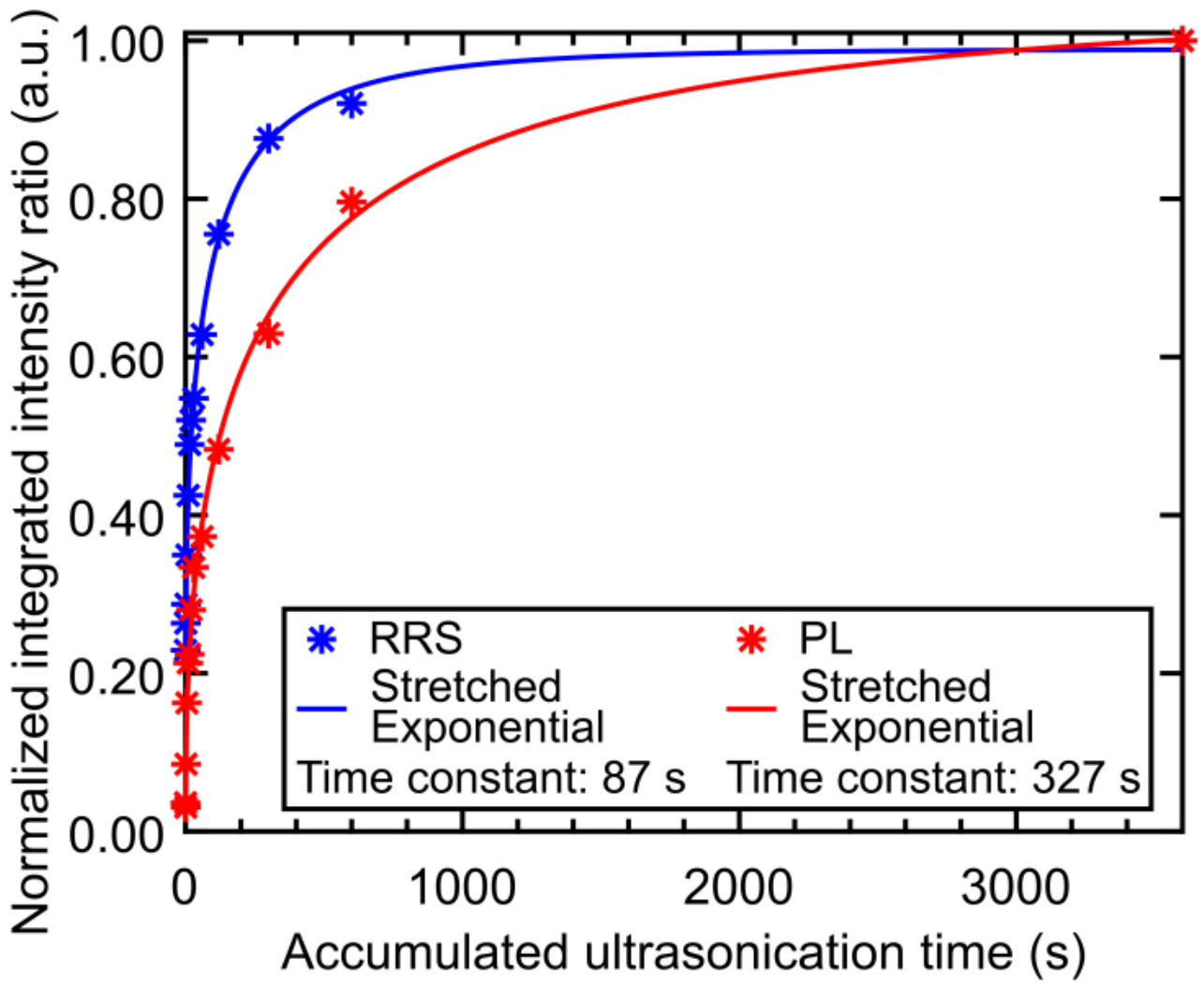

**Figure 5.** Normalized ultrasonication time-dependent intensity ratios of the signals from extracted SWCNTs with respect to the signals of the remaining DWCNTs measured by RRS (blue) and PL (red) spectroscopy. The SW- and DWCNT intensities are calculated by integrating the respective signals over the ranges indicated by the dashed lines and highlight colors in Figure 4 (and Figure S7 in the SI) and by averaging over all excitation wavelengths for RRS or PL separately. The data are fitted using a stretched exponential with time constants $\tau$ of 87 and 327 s and exponents $s$ of 0.52 and 0.47 for respectively RRS and PL using the equation: $r(t) = f - b\, e^{-(t/\tau)^s}$.

In both RRS and PL, however, we clearly prove that even brief ultrasonication times of only several seconds already drastically destroy the DWCNTs resulting in a significant fraction of SWCNTs in the sample, and approximately 50% of the total amount of extracted SWCNTs is obtained after only 60 s. Moreover, a saturation is clearly observed, reaching after one hour a total drop in DWCNT inner shell RBM intensity to approximately 40%, whereas the DWCNT outer shell PL drops to only 30% of its initial intensity (see Figure S10 in the SI). After this saturation is achieved, the CNTs in the sample are 1.6 times more defective compared to before the extraction process according to the integrated D/G-ratio measured at 725 nm (see Figure S11 in the SI).

Despite this rapid extraction, the saturation suggests that not all DWCNTs can be destroyed during the process. This could hint at an extraction mechanism where the DWCNT outer wall is required to be amply defected to be able to break and to open after which the inner shell can slide out. On the other hand, the extraction could be more efficient for those inner and outer tube combinations where the van der Waals interaction is minimal, while others might remain intact [16]. This might explain why some statistical HRTEM studies that applied ultrasonication conclude that not all inner and outer tube combinations are possible [16], whereas other studies claim there is no chiral correlation between the two shells [46,47]. Another possible explanation for this saturation effect could be related to the DWCNT length. Only in case of sufficiently long DWCNTs will the inner and outer shells be pulled apart, while the ultrasonication effect will lose its grip on shortened DWCNTs. Nevertheless, at the same time it can be expected that longer DWCNTs exhibit a stronger van der Waals interaction between the inner and outer shells. Finally, these ultrasonication time-dependent measurements prove that ultrasonication alone is sufficient to extract the inner tubes from the DWCNTs unlike the work of Rohringer *et al.* [34] where it was suggested that DGU is required for the extraction.

According to the extraction mechanism proposed by Miyata *et al.* [36], both the inner and outer shell of the DWCNT should lead to respectively small- and large-diameter SWCNTs. The RRS and PL spectra in Figure 4 indeed confirm the build-up of the extracted inner shell SWCNTs, but show no clear increase in outer shell SWCNT signals. There are multiple reasons why these outer shell SWCNTs might be missed with optical spectroscopy. First, it is expected that for such large-diameter SWCNTs their Raman and PL signals strongly overlap with the remaining DWCNT outer shell signals, complicating their direct assessment [14,15]. The latter is not the case for the inner tube diameter range, due to the strong shifted electronic transitions of inner CNTs of DWCNTs with respect to SWCNTs [43,44]. Second, the Raman cross sections and PL quantum efficiencies are known to be smaller for such large-diameter SWCNTs in comparison with the thinner extracted

SWCNTs [41,48]. In addition, the RBMs of these big SWCNTs strongly overlap with each other, resulting in a broad peak, which complicates the differentiation between the signals of the extracted outer shell SWCNTs and the remaining DWCNT outer shells even further. Third, the outer shell SWCNT are opened during the process, causing them to be water-filled and hence drastically quenched in fluorescence and in Raman spectroscopy [41,49]. Furthermore, since the fluorescence of CNTs is strongly quenched at their ends or at local defect sites along the backbone of the CNT [50], ultrasonication, which cuts CNTs in shorter pieces and creates defects, will inevitably result in relatively lower PL quantum efficiencies.

The PL of the DWCNT outer shells, measured at 1050 nm in Figure 4, decreases steadily as a function of ultrasonication time. Hence, no increasingly stronger, overlapping PL of the outer shell SWCNTs is detected during ultrasonication, most probably because of the aforementioned quenching possibilities. In case of RRS, however, the DWCNT outer shell RBMs, centered at 170 and 160 $cm^{-1}$ for respectively 725 and 780 nm as presented in Figure 4, show a differently decreasing behavior as a function of ultrasonication compared to that of the DWCNT inner shell RBMs. Again, to exclude any alignment variations, the intensity ratio between the DWCNT inner and outer tube RBM signals is calculated as a function of ultrasonication time by integrating their respective RRS signals and adding the contributions for both excitation wavelengths. Figure 6 shows how this ratio decreases as a function of ultrasonication time. However, both the DWCNT inner and outer tube RBMs are expected to diminish simultaneously in intensity when the shells of a DWCNT are pulled apart by ultrasonication. The slower decrease rate of the DWCNT outer shell RBMs compared to that of the DWCNT inner shell RBMs could be explained by an increasing contribution of the overlapping outer shell SWCNT RBMs. Hence, we prove that both the DWCNT inner and outer shells indeed form respectively small-and large-diameter SWCNTs, although these outer shell SWCNTs are expected to be heavily quenched in PL due to the possible water-filling or ultrasonication-induced defects.

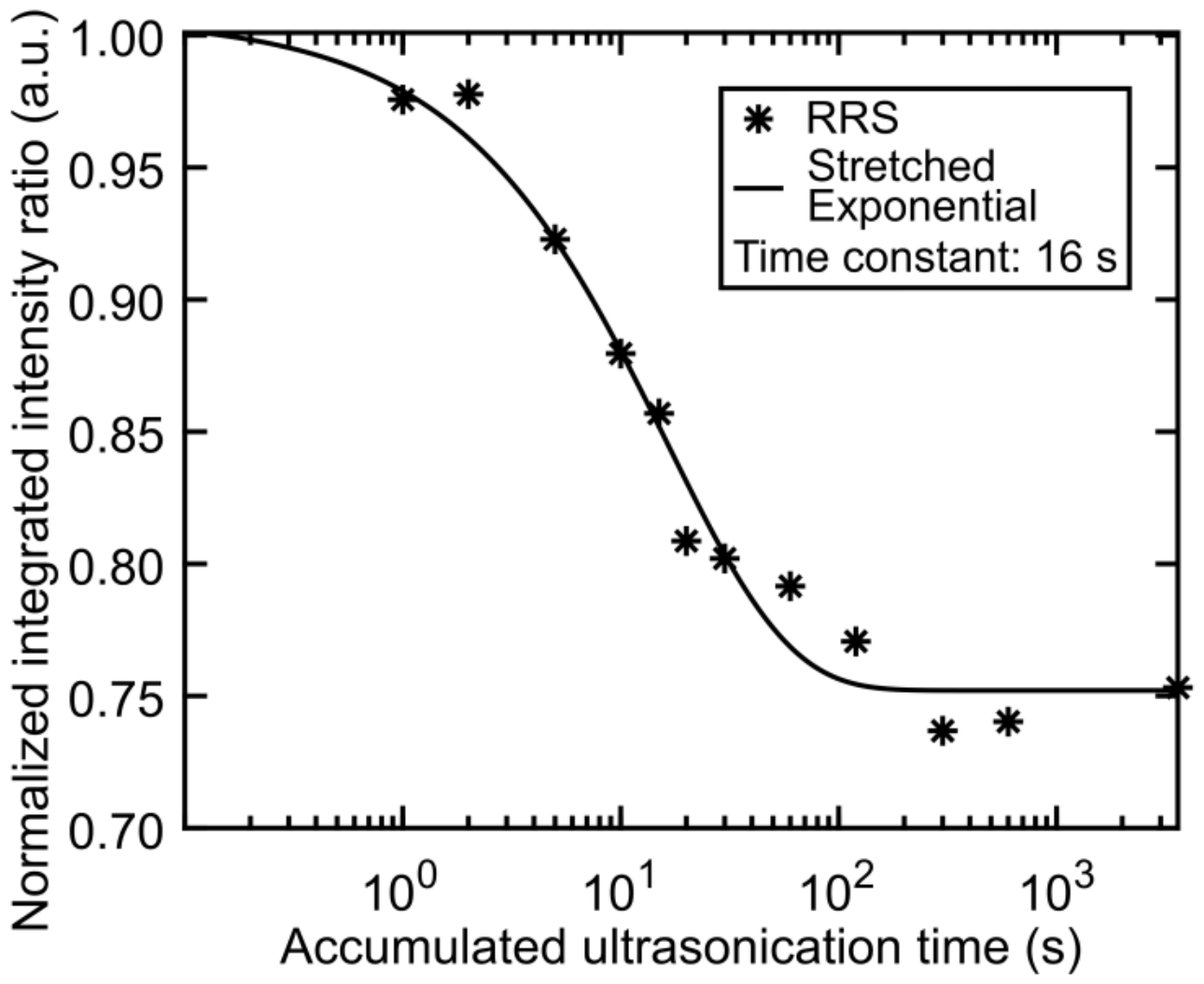

**Figure 6.** Decreasing ratio of the DWCNT inner to outer tube RBM intensities as a function of ultrasonication time. The ratio, calculated by integrating the respective DWCNT inner and outer tube RBM signals over fixed intervals and averaging the contributions over both excitation wavelengths, follows a stretched exponential curve with a shorter time constant compared to the ratio between the SW- and DWCNT inner shell RBMs. In view of the extraction mechanism the DWCNT inner and outer shell RBM intensities are expected to diminish at the same rate, resulting in a constant ratio. The detected decreasing ratio could be explained by an increasing contribution of the overlapping RBMs of the ultrasonication-induced extracted outer shell SWCNTs.

In addition to the detailed PL and RRS ultrasonication-time-dependent studies, the ultrasonication-induced inner shell extraction, resulting in SWCNTs, is also clearly visible form absorption spectroscopy. This is illustrated in Figure 7, where the absorption spectrum of the non-sonicated purified DWCNT sample is compared with that of the one hour long ultrasonicated DWCNTs. To facilitate the comparison, the Mie-scattering background was removed from these spectra, see Figure S12 in the SI, after which these spectra were normalized. As expected, new absorption features clearly arise after ultrasonication, especially in the small-diameter wavelength range of 600 nm to 1400 nm, which are attributed to the extracted inner shells of the DWCNTs.

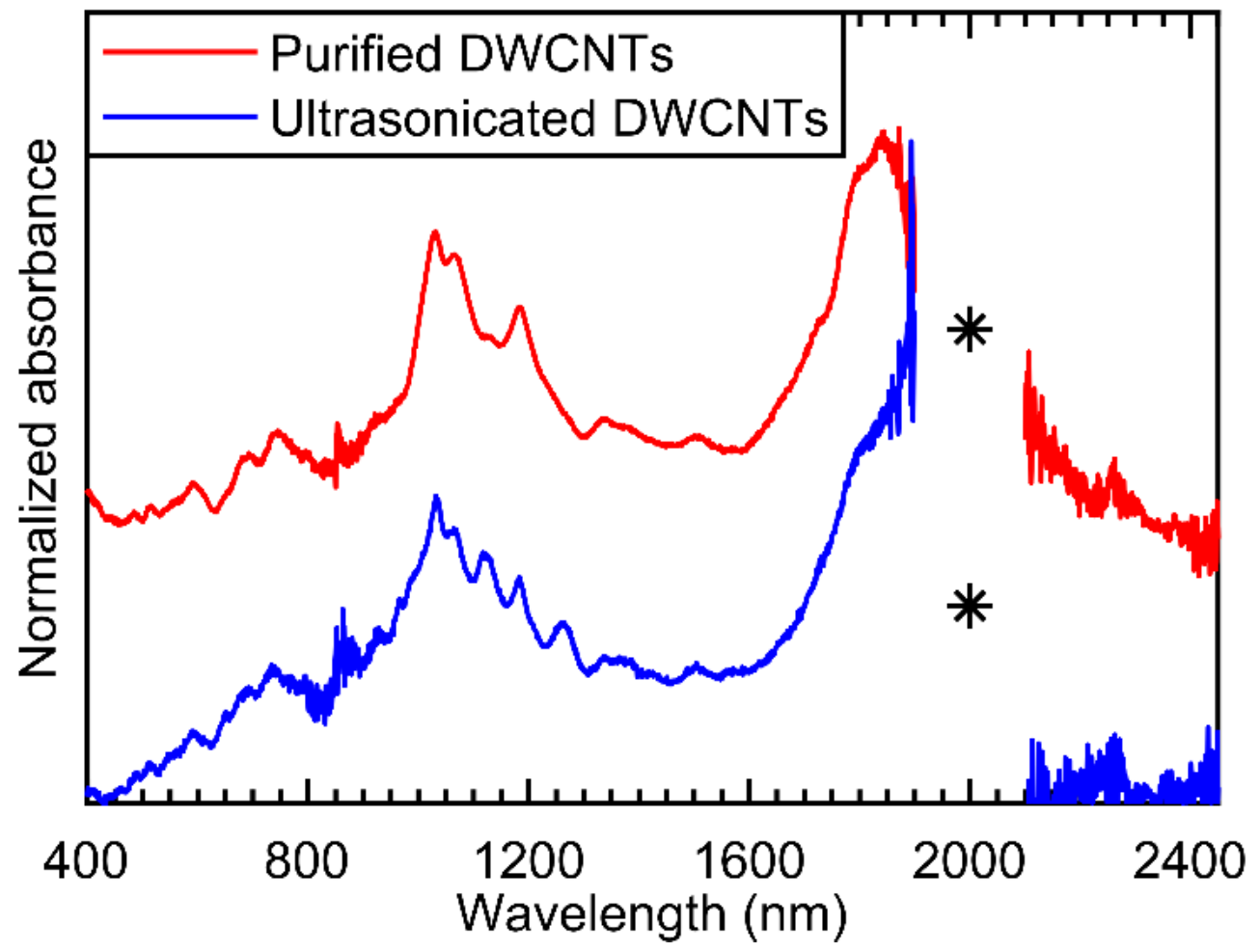

**Figure 7.** Comparison of the normalized absorption spectra of the purified DWCNTs before (red trace) and after extensive ultrasonication (blue trace) for one hour. The spectra's background have been removed (see Figure S12 in the SI) in order to ease the comparison. New absorption features are clearly visible in the ultrasonicated DWCNT sample that are attributed to the extracted inner shell SWCNTs. The asterisks indicate where the strong $D_2O$ absorption band complicates the background subtraction.

### 3.4 Extracted inner shell separation

After saturating the extraction by ultrasonicating the sample for one hour, making sure that the maximal amount of SWCNTs is extracted, we apply a final DGU step to separate these extracted SWCNTs from the remaining DWCNTs. If successful, this proves that the SWCNTs are completely extracted from the DWCNTs and, hence, are individually solubilized. Indeed, only if both shells are fully extracted from each other will this DGU separation lead to SW- and DWCNTs with density distributions similar to the distributions of the pristine SW- and DWCNTs from the earlier purification DGU runs shown in Figure 2 and Figure 3, where both CNT species are separated from each other based on their differences in density. To compare, we likewise sort the non-sonicated, purified DWCNT reference sample in the same DGU run. Since this reference sample has not been ultrasonicated, no SWCNTs are extracted and, hence, only DWCNTs are expected to be present after DGU sorting. If so, this confirms once more that DGU itself does not cause any inner shell extraction.

To characterize the SW- and DWCNT distributions inside the ultracentrifuge tube we again use *in situ* RRS and PL spectroscopy at excitation wavelengths similar to the ones from previous DGU separations. A selection of these *in situ* measurements, respectively measured at 725 and 780 nm for RRS, and 650 and 1050 nm for PL, is shown in Figure 8 (see Figure S14 and Figure S15 in the SI for additional excitation wavelengths).

The RRS and PL measurements of the purified DWCNT reference sample, displayed in the top panels of Figure 8, only show the characteristic DWCNT features, as expected from the *prior* two-step DGU purification of the DWCNT sample. The lack of any SWCNT signal proves that DGU itself does not extract any SWCNTs from DWCNTs, in contrast to the work of Rohringer *et al.* [34]. The DWCNT outer shell RBMs can be distinguished on top of the NycoDenz-peak, while the DWCNT inner shell signals at larger Raman shifts can be identified as indicated by the chiral indices on the color maps. In addition, the unexpectedly strong DWCNT outer shell fluorescence is present as well. As is expected, all of the DWCNT signals of this unaltered reference sample occur at similar depths (hence densities) in the ultracentrifuge tube (22 – 27 mm) compared to the results obtained in previous DGU runs presented in Figure 2 and Figure 3 (see also SI Section 3 and Figure S14 and Figure S15).

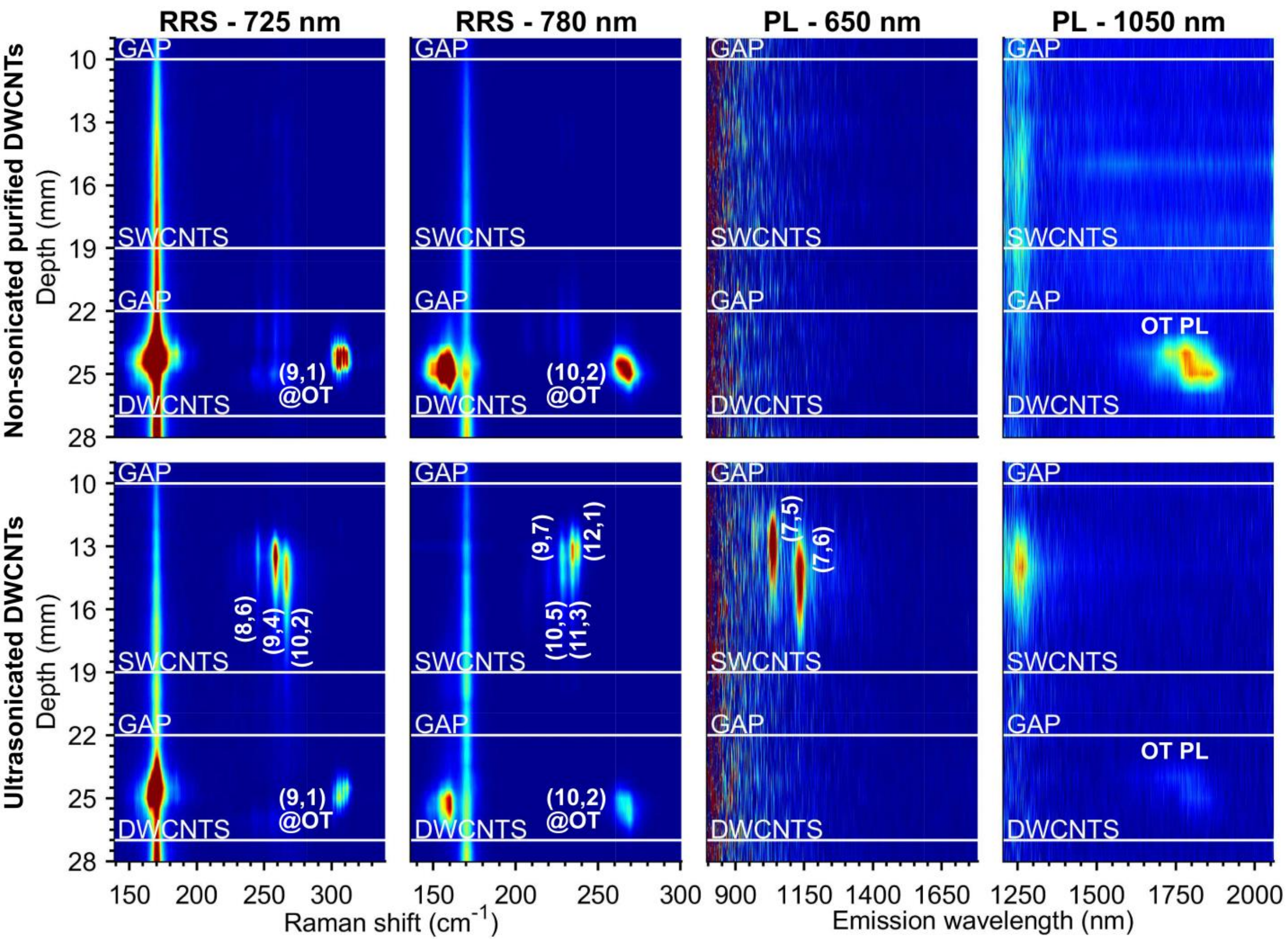

**Figure 8.** *In situ* RRS and PL spectra as a function of depth in the ultracentrifuge tube (see Figure S13 in the SI for the density-to-depth relation) for the non-sonicated purified DWCNT reference sample (top panels) and the one-hour long ultrasonicated DWCNTs (bottom panels). The RRS color maps, measured at 725 and 780 nm, show for both samples the typical DWCNT inner and outer shell RBMs at the bottom of the ultracentrifuge tube around 22 – 27 mm (higher density). Only in case of the ultrasonicated DWCNTs additional signals are detected at the top around 10 – 19 mm (lower density) that are assigned to the extracted inner shell SWCNTs. The PL color maps, measured at 650 and 1050 nm, are all plotted on the same color scale. They analogously show fluorescence of the DWCNT outer shells at similar depths as the RRS color maps, while only in the ultrasonicated DWCNT sample PL of the extracted inner shell SWCNTs is measured at the top, similar again to the RRS color maps. In both RRS and PL, the DWCNT signals in the ultrasonicated sample are much weaker than those in the reference sample. Both CNT distributions are physically separated from each other, proving hereby that the SWCNTs are fully extracted from the DWCNTs. The assigned fractions and chiralities are indicated on the color maps in white.

The *in situ* RRS and PL measurements on the extensively ultrasonicated DWCNT sample are displayed in the bottom panels of Figure 8. Identical to the reference sample, DWCNT RRS and PL signals are detected. These DWCNT signals are, however, much weaker than in the non-sonicated purified DWCNT sample since more than 60% of the DWCNTs are destroyed during the ultrasonication-induced extraction as mentioned earlier (see Figure S10 in the SI). Despite being less abundant due to the ultrasonication, the remaining DWCNT signals still occur at depths (densities) similar to the ones in the reference sample, confirming that the buoyant density of the remaining DWCNTs is unaffected by the extensive ultrasonication (see SI Section 3).

Besides the leftover DWCNT signals, the ultrasonicated sample has additional signals near the top of the ultracentrifuge tube (10 – 19 mm) that arise from the extracted inner shell SWCNTs. Indeed, these signals match perfectly in both Raman shift and emission wavelength with the increasing extracted SWCNT signals from Figure 4. Furthermore, their positions are in agreement with those of water-filled SWCNTs, which is expected for such ultrasonicated and thus opened SWCNTs [38,49]. The chiralities of these extracted inner shell SWCNTs are assigned similarly to the indexation of the as-synthesized SWCNTs removed from the DWCNTs during the first DGU run, see Figure 2, and overlaid in white on the color maps. In comparison with the SWCNTs from the first DGU, however, the extracted inner shell SWCNTs appear to contain more small diameters, which might be explained by the fact that the inner shells are limited in diameter by the DWCNT outer shell. Furthermore, this difference in diameter distribution confirms that the SWCNTs in the parent sample are indeed most likely co-produced during the CCVD synthesis instead of extracted SWCNTs.

The extracted inner shell SWCNTs not only have matching spectra compared to the as-synthesized SWCNTs with the same chirality, but they occur as well at similar depths in the ultracentrifuge tube, hence at similar densities (see Figure S14 and Figure S15 in the SI). This proves the complete extraction of individual shells from the DWCNTs. Furthermore, no other signal is detected in between the extracted SWCNTs and leftover DWCNTs, confirming indeed that there are no partially extracted SW-DWCNT structures. However, despite the nicely resolved extracted inner shell SWCNT features, no outer shell SWCNT RBM nor fluorescence is detected, possibly because of the overlapping NycoDenz Raman mode and filling- or defect-induced PL quenching of these large-diameter SWCNTs, respectively, as discussed in Section 3.3.

### 3.5 Transmission electron microscopy characterization

The detection of outer shell SWCNTs has been troublesome throughout the various optical spectroscopy measurements because of overlapping signals from the DWCNT outer shells and the NycoDenz Raman mode. Nevertheless, we could indirectly, yet successfully prove that the amount of outer shell SWCNTs increases during ultrasonication, see Figure 6. To unambiguously verify that ultrasonication indeed extracts the DWCNT inner shell and, at the same time, forms an outer shell SWCNT, instead of destroying the outer shell, we performed detailed statistical HRTEM characterization studies on the purified and ultrasonicated DWCNT samples.

Although HRTEM is a powerful technique to estimate statistically the diameter distribution of a CNT sample, it can be subject to many biases and thus a well-chosen methodology that avoids these biases is required [51]. Generally, porous supporting grids are employed in TEM to deposit the CNTs on. The effective porosity of these supporting grids can dramatically affect the observed populations since CNTs are easily flushed through these pores while rinsing the surfactant molecules off the grid. In this study one of the DWCNT samples has been extensively ultrasonicated on purpose, hence we expect the CNTs to be drastically shortened [27,28,38], which makes it difficult to preserve them on the grid during preparation. Therefore, specific supporting grids are used instead, where an additional graphene layer on top of lacey carbon maximizes the supporting area (see Figure S16 in the SI). These grids effectively prevent the CNTs from being flushed away, as is exemplified in Section 4 of the SI and in Figure S17. Moreover, the additional graphene layer creates a weak continuous contrast with respect to the CNTs, allowing for a clear determination of the number of walls and their respective diameters.

From the HRTEM images we determine the diameter distributions and abundances of the various CNT species present in the ultrasonicated DWCNT sample and compare them with those present in the purified DWCNT sample. Contrarily to HRTEM, where despite the improved TEM grids the bias is typically towards larger diameter CNTs [51], in our optical spectroscopy measurements only CNTs with a diameter up to 1.8 nm are detected. Hence, to compare the statistical results of HRTEM, shown in Figure 9, with the earlier spectroscopic conclusions, the statistical data is deliberately limited to those CNTs that have an inner shell-diameter below 1.8 nm. This cut-off limit is indicated in Figure 9 by a dashed vertical black line overlaid on the diameter distributions of the DWCNT inner shells and SWCNTs and by visualizing the larger diameter CNTs with transparent bars. Only those CNTs with an inner diameter below this limit are considered when calculating the abundances and their relative ratios to obtain a better comparison with the optical spectroscopic results. The top and bottom panels in Figure 9 show, from left to right, the diameter distributions of the DW- and SWCNTs and their total abundances for the purified and ultrasonicated DWCNT samples, respectively.

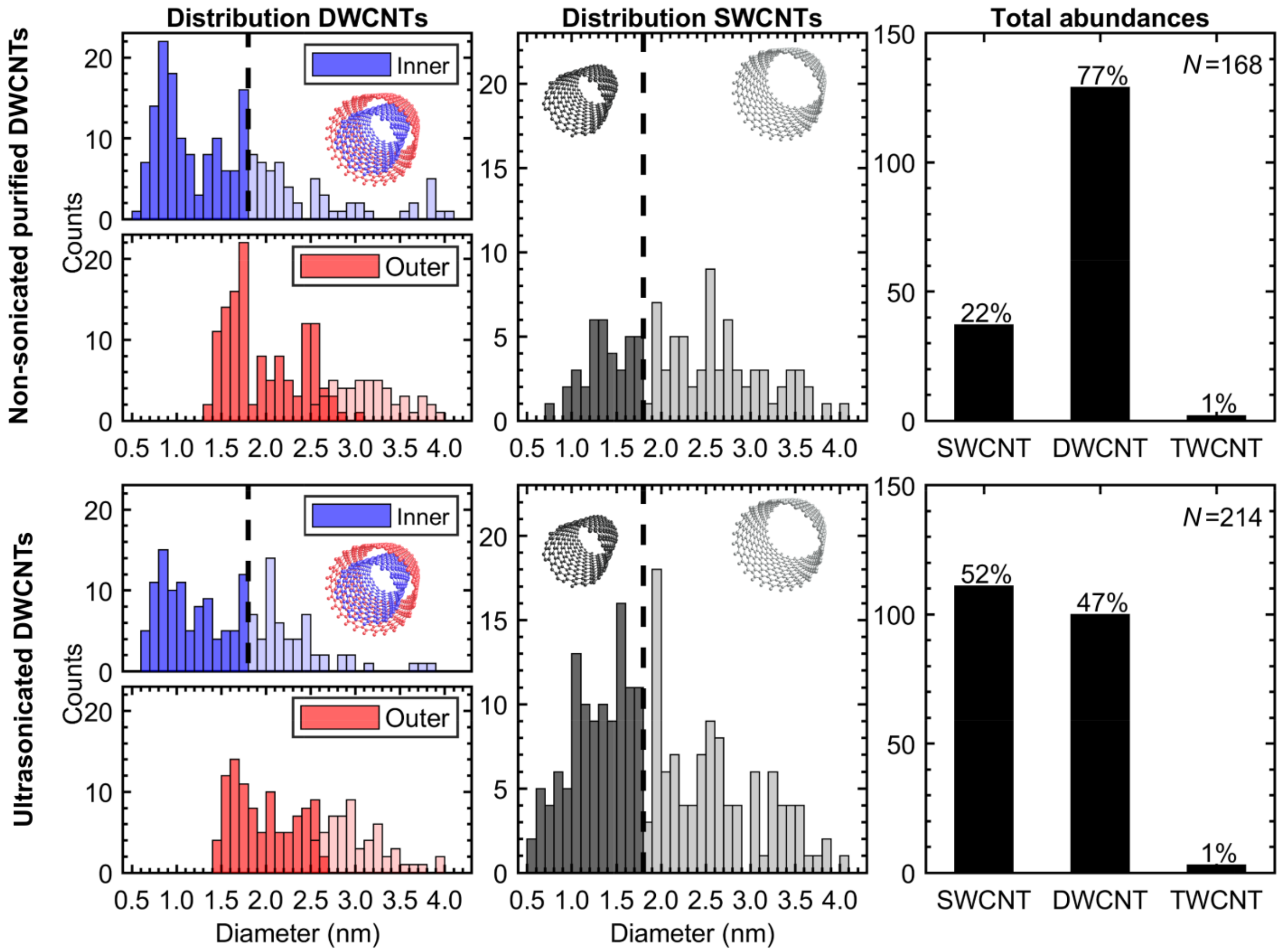

**Figure 9.** Comparison of the statistical HRTEM data for the DWCNT samples before (top panels) and after extensive ultrasonication (bottom panels). From left to right, the panels show the diameter distributions of the DW- and SWCNTs and their total abundances. To compare the HRTEM results with the spectroscopic data, the distributions and following abundances are limited to those CNTs that have an inner diameter below 1.8 nm, as visualized on the diameter distributions of the DWCNT inner shells and SWCNTs by the dashed vertical black line. CNTs with larger inner shell-diameters have transparent bars. *N* denotes the total number of observed CNTs with inner diameter below 1.8 nm. After ultrasonication, there are significantly more SWCNTs than in the purified DWCNT sample, especially below 1.8 nm, in agreement with the earlier spectroscopic results.

As expected from the rigorous DGU purifications, the purified DWCNT sample consists mostly of DWCNTs. Only 22% of all 192 observed CNTs below the cut-off limit are SWCNTs. These are most probably the kinked/defective SWCNTs that persisted in the solution despite the two consecutive DGU runs. This amount of SWCNTs is in stark contrast with the ultrasonicated

DWCNT sample, where significantly more SWCNTs, 52%, have been observed than in the purified DWCNT sample (22%). This drastic change in ratio between the abundances of the SW- and DWCNTs before and after ultrasonication is in perfect agreement with our earlier spectroscopic conclusions. Hence, statistical HRTEM analysis confirms that ultrasonication successfully extracts the inner shell from the outer tube of individually solubilized DWCNTs to form SWCNTs.

The abundances of the DWCNT inner and outer shells shown in Figure 9 are diminished equally after ultrasonication without that their diameter distributions undergo any significant change in the mean or standard deviation. In view of the proposed extraction mechanism where the inner shell slides out of the outer tube, it is indeed sensible that both abundances decrease equally. The fact that their diameter distributions remain the same indicates that the extraction mechanism shows no pronounced diameter-dependence. At the same time, this mechanism implies that not only the number of small-diameter inner shell SWCNTs should increase but also the amount of larger-diameter outer shell SWCNTs. Indeed, the abundance of the SWCNTs in Figure 9 increases after ultrasonication for all diameters. This confirms, in addition to the inner shell SWCNTs, the presence of the earlier detected outer shell SWCNTs, now through a direct observation. However, these extracted SWCNTs are often heavily truncated and damaged, as illustrated by representative HRTEM images of both samples in Figure S18 and Figure S19 in the SI. This is especially true for the larger diameter SWCNTs, which is sensible since they originate from the DWCNT outer shells and thus are in direct contact with the ultrasonic waves while shielding the inner shell [4]. In addition to the defects, the truncation of these SWCNTs inevitably fill them with water [38], possibly explaining why the outer shell SWCNTs are so hard to observe with optical spectroscopy. Besides an increase in outer shell SWCNTs, there are significantly more small-diameter SWCNTs observed after ultrasonication, especially within the diameter range below the cut-off limit of 1.8 nm where diameters as thin as 0.6 nm are detected. When taking into account that such small diameters are easily overlooked in HRTEM or even destroyed by the electron beam during image acquisition [51], this is even more impressive. Compared to normal SWCNT samples, such thin shells are more stable in DWCNTs because they are shielded from the harsh outer environment during growth by the outer tube [4,5]. Hence, the ultrasonication-induced extraction mechanism unraveled in this work can be purposely used to extract these specific small-diameter SWCNTs from DWCNTs, as has already been demonstrated in the analysis of inner tube chiralities grown by thermal conversion of fullerene-like precursors encapsulated in outer tubes [52].

Finally, the fact that most of the observed outer shell SWCNTs were heavily damaged, while at the same time many DWCNTs were observed, combined with the quickly saturating extraction, hints at an extraction mechanism that possibly depends on how defective the DWCNTs are. Indeed, the more defects the outer shell of a DWCNT initially has, the more easily it will break in two during ultrasonication followed by the inner shell extraction. On the other hand, defect-free DWCNTs require more time before extraction can occur, if at all, in line with the slowing down extraction shown in Figure 5. The remaining 40% of DWCNTs thus could possibly be those that were hardly defected in the beginning. In addition, as has been discussed in section 3.2, other factors such as the DWCNT length and even the inter-layer coupling strength could significantly influence how easily the two layers of a particular DWCNT can be pulled apart during ultrasonication and can thus affect the extraction mechanism, all effects that could be investigated in the future.

## 4. Conclusion

In this work, spectroscopically purified DWCNTs have been used to characterize in detail the ultrasonication-induced extraction of SWCNTs. By ultrasonicating these purified DWCNTs and characterizing the sample with RRS and PL spectroscopy, we not only unambiguously proved the facile inner shell extraction spectroscopically, but also showed that ultrasonication alone is sufficient for the extraction unlike suggested in reference [34] where DGU is needed as well. Moreover, by ultrasonicating in subsequent time steps and spectroscopically probing the sample intermittently, we unraveled the quickly saturating time dependency of the extraction and demonstrated that even brief ultrasonication durations shorter than one minute are sufficient to reach ~50% of the total SWCNT extraction, which then completely dominates the fluorescence and RBM signals in the sample. After one hour of ultrasonication, approximately 60% of the initial DWCNTs are destroyed according to our spectroscopic results. After extensive ultrasonication we separated the extracted SWCNTs from the remaining DWCNTs *via* DGU, hereby proving that the newly formed SWCNTs are individually solubilized. Finally, we employed HRTEM to determine the abundances and diameter distributions of the various CNT species in the purified and extensively ultrasonicated DWCNT sample. These results revealed, in addition to the formation of inner shell SWCNTs, the presence of large-diameter SWCNTs formed from the outer shells of DWCNTs. Furthermore, the observation of both intact DWCNTs as well as heavily damaged outer shell SWCNTs with HRTEM suggested that the extraction mechanism depends possibly on how defective the outer shell of the DWCNT is. In short, we unquestionably prove that ultrasonication has devastating effects on individually solubilized DWCNTs resulting in new, strongly fluorescing SWCNTs that, among other, can easily complicate the quantification of the DWCNT inner shell's fluorescence quantum yield.

**Supporting Information Available:** The Supporting Information contains details and additional figures on the preparation and purification of the DWCNT sample as well as the *in situ* spectroscopic characterizations, additional PL spectra and more detailed fits of the ultrasonication time-dependent intensity ratios, additional details on and spectroscopic results of the DGU separation of the extracted DWCNTs, relevant information on the sample preparation for the HRTEM measurements and the effect of the type of TEM grids used on the diameter distributions.

**Corresponding Authors:**

* Correspondence should be addressed to Sofie Cambré or Wim Wenseleers of the University of Antwerp.
E-mail addresses: sofie.cambre@uantwerpen.be and wim.wenseleers@uantwerpen.be

**Notes:**

The authors declare no competing financial interest.

**Orcid IDs:**

- Maksiem Erkens: 0000-0001-9394-7655
- Sofie Cambré: 0000-0001-7471-7678
- Emmanuel Flahaut: 0000-0001-8344-6902
- Frédéric Fossard: 0000-0002-7886-5309
- Annick Loiseau: 0000-0002-1042-5876

- Wim Wenseleers: 0000-0002-3509-0945

**Funding Sources:**

- PhD fellowship from the Research Foundation of Flanders (FWO) for M.E. (Grant number: 11C9220N)

- Starting grant of the European Research Council (ERC) for S.C. (Grant number: 679841)

- FWO projects with grant numbers G040011N, G02112N, G035918N and G036618N

**Acknowledgments:**

This work was financially supported by a PhD fellowship for M. Erkens of the Research Foundation of Flanders (FWO, grant number: 11C9220N), several FWO projects (G040011N, G02112N, G035918N and G036618N). M. Erkens and S. Cambré also acknowledge funding from the European Research Council (ERC) starting grant of S. Cambré (Grant number: 679841).

**Graphical Abstract:**

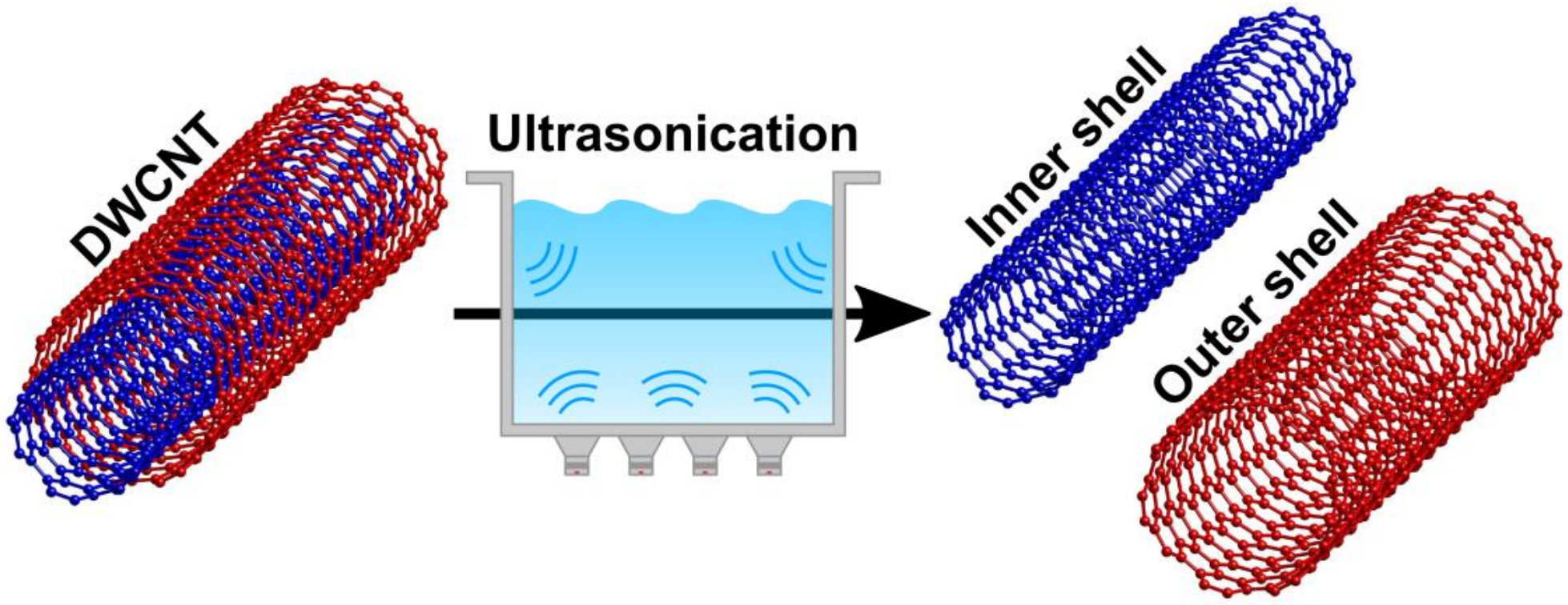

# Supporting Information for:

# Ultrasonication-Induced Extraction of Inner Shells from Double-Wall Carbon Nanotubes Characterized via In Situ Spectroscopy after Density Gradient Ultracentrifugation.

*Maksiem Erkens [1], Sofie Cambré [1,*], Emmanuel Flahaut [2], Frédéric Fossard [3], Annick Loiseau [3], and Wim Wenseleers [1,*]*

[1] Nanostructured and Organic Optical and Electronic Materials (NANOrOPT), Department of Physics, University of Antwerp, B-2610 Antwerp, Belgium
[2] CIRIMAT, Université de Toulouse, CNRS, INPT, UPS, UMR CNRS-UPS-INP N°5085, Université Toulouse 3 Paul Sabatier, Bât. CIRIMAT, 118, route de Narbonne, 31062 Toulouse cedex 9, France
[3] Laboratoire d'Etude des Microstructures, CNRS-ONERA, Université Paris-Saclay, Châtillon, France

* Correspondence should be addressed to: sofie.cambre@uantwerpen.be or wim.wenseleers@uantwerp.be

## Table of Contents

## *In situ* spectroscopic characterization of the DWCNT purification

### Depth-to-density relation

After carefully solubilizing the CCVD-grown DWCNT powder in a 2 wt./V% DOC/$D_2O$ solution and subsequently centrifuging any undissolved species out of the colloidal solution, the DWCNTs are further purified from as-synthesized SWCNTs and bundles *via* DGU. The exact densities of the top- and bottom-phase in the ultracentrifuge tube are determined empirically to optimally separate the DWCNTs from the SWCNTs (lower density) and bundles (higher density) within the physical height range of the ultracentrifuge tube.

Once isopycnic equilibrium is reached, the density profile in the ultracentrifuge tubes is determined by using the analysis protocol described in reference [1]. *Via* absorption spectroscopy the density is obtained at discrete depths in one of the ultracentrifuge tubes. These points are then interpolated using a piecewise cubic Hermite polynomial. The resulting density-to-depth relation acquired for the first DGU run is displayed in Figure S1, next to a picture of such an ultracentrifuge tube after DGU. Near the bottom of the ultracentrifuge tube, the density rises quickly and, ideally, this non-linear range is designated for the CNT bundles. Therefore, the densities are chosen so that the CNT distributions of interest mostly fall within the more linear part of the probed depth range during the *in situ* measurements. The probed depth range in the *in situ* measurements is indicated by the horizontal dashed lines in Figure S1. With this relation, the following *in situ* RRS and PL spectra can be directly plotted as a function of density instead of depth, as is illustrated in the following section and Figure S2.

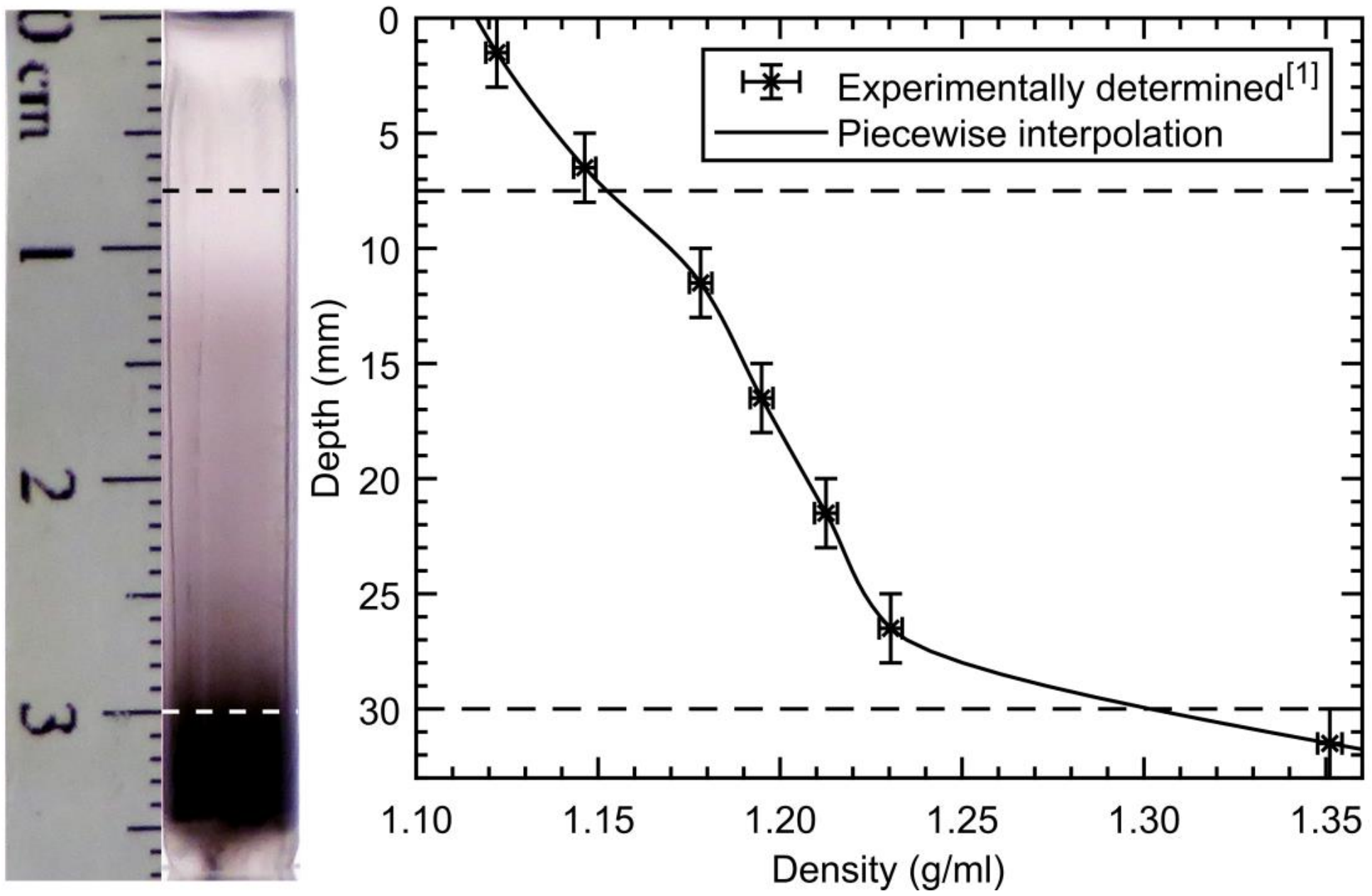

**Figure S1.** Density-to-depth relation inside the ultracentrifuge tubes after the CNTs reach isopycnic equilibrium with the gradient medium for the first DGU purification run, determined by examining one ultracentrifuge tube as described in reference [1]. The discrete data points are interpolated with a piecewise cubic Hermite polynomial. Throughout the depth of the ultracentrifuge tube, the density increases steadily and only near the bottom, where all the bundles are aggregated, the density rises rapidly. The SW- (7.5 – 16.5 mm) and DWCNT (21.5 – 25 mm) distributions of interest are located well within the more linear range. The horizontal dashed lines indicate the probed depth range of the *in situ* measurements.

**Reproducibility of the gradient**

Although, according to reference [1], the acquired density profile is equal for all of the ultracentrifuge tubes from the same DGU run, we first quickly verify if the other remaining three ultracentrifuge tubes have the same CNT distributions at indeed similar depths and densities by measuring *in situ* RRS at 725 nm on all three ultracentrifuge tubes. This allows us either to confirm the identical density-to-depth relation for all the ultracentrifuge tubes, or to adjust for any inconsistencies. In case of the first DGU separation, these quickly measured *in situ* RRS spectra, plotted as a function of both depth and density in Figure S2, show a good similarity in CNT distributions between the remaining three different ultracentrifuge tubes of the same DGU run. Furthermore, Figure S2 compares the same *in situ* RRS maps plotted as a function of depth and density, using the relation established in Figure S1, to illustrate how the various signals distributed along the depth in the ultracentrifuge tube are translated to density. Below a density of 1.25 g/ml no signals are detected as this region was designated for the collection of any bundles in the sample.

***In situ* characterization**

Next, one of the ultracentrifuge tubes is characterized in detail using both *in situ* RRS, with excitation wavelengths 670, 695, 725 and 780 nm, as well as *in situ* PL spectroscopy, at excitation 570, 650, 730, 800, 860, 930, 970, 1050, and 1130 nm. The idea is to precisely monitor the specific chirality distributions of all CNT species present as a function of density. The resulting *in situ* RRS and PL color maps for the first DGU run are summarized in Figure S3 and Figure S4 respectively.

Thanks to this thorough *in situ* characterization we can notice how for various excitation wavelengths the SW- and DWCNTs are distributed differently as each excitation wavelength probes a different set of chiralities that each have a slightly different density. The assigned chiralities are indicated by their chiral indices in white on the color maps. Based upon this information we can assign the best fractions, indicated by the white horizontal lines on the color maps of Figure S3 and Figure S4. We deliberately chose to leave gaps in between the important SWCNT and DWCNT fractions, to minimize overlap with other CNT species at the cost of losing some CNTs.

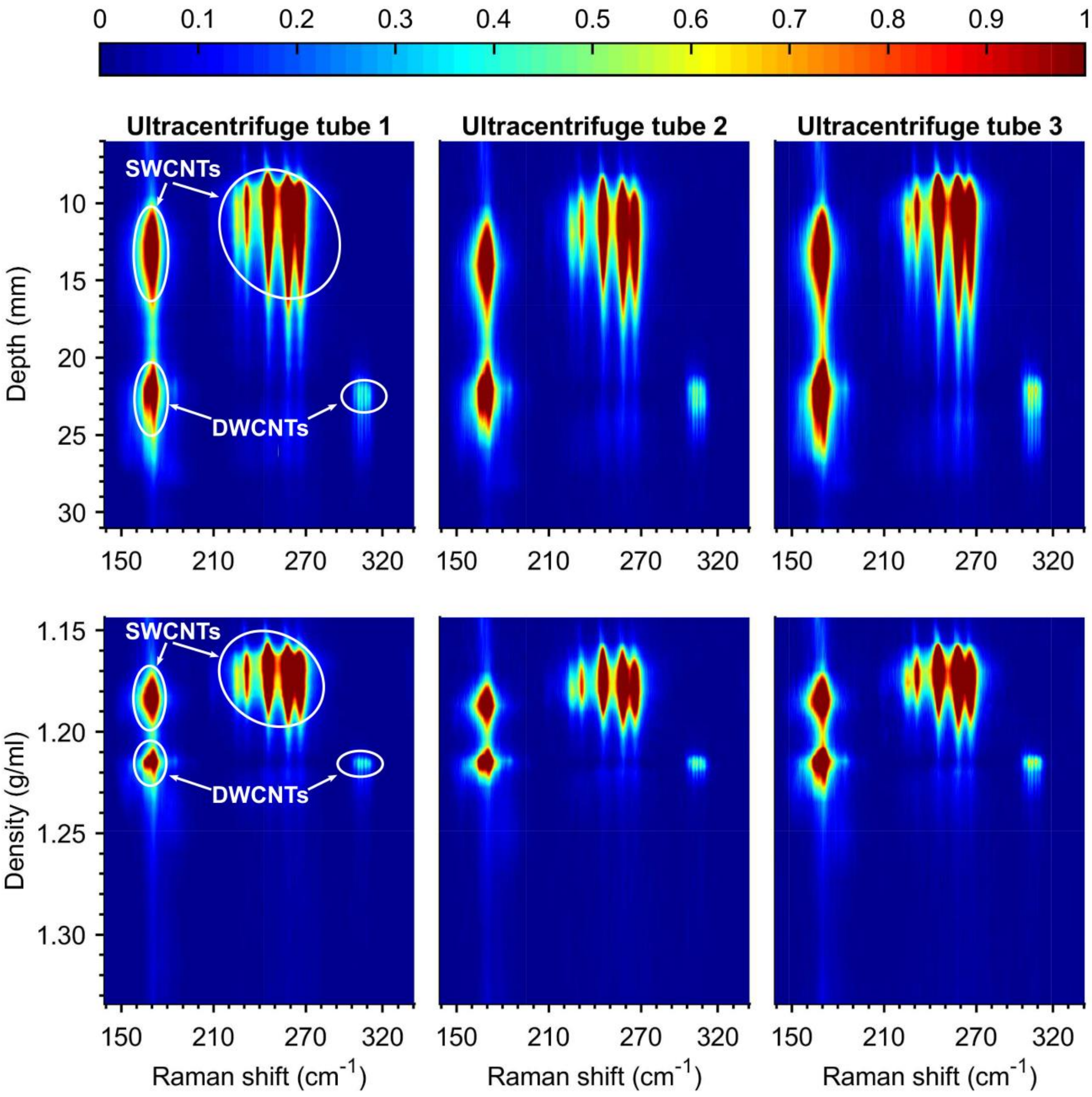

**Figure S2.** Color maps of the quick *in situ* RRS characterization at excitation 725 nm, where the Raman shift is plotted as a function of density in each of the three ultracentrifuge tubes. The color maps of these three different ultracentrifuge tubes from the first DGU run show a great similarity in both SW- as well as DWCNT distribution, verifying thus the reproducibility of the gradient. In addition, these signals are well separated from the highest densities where the bundles are expected to accumulate.

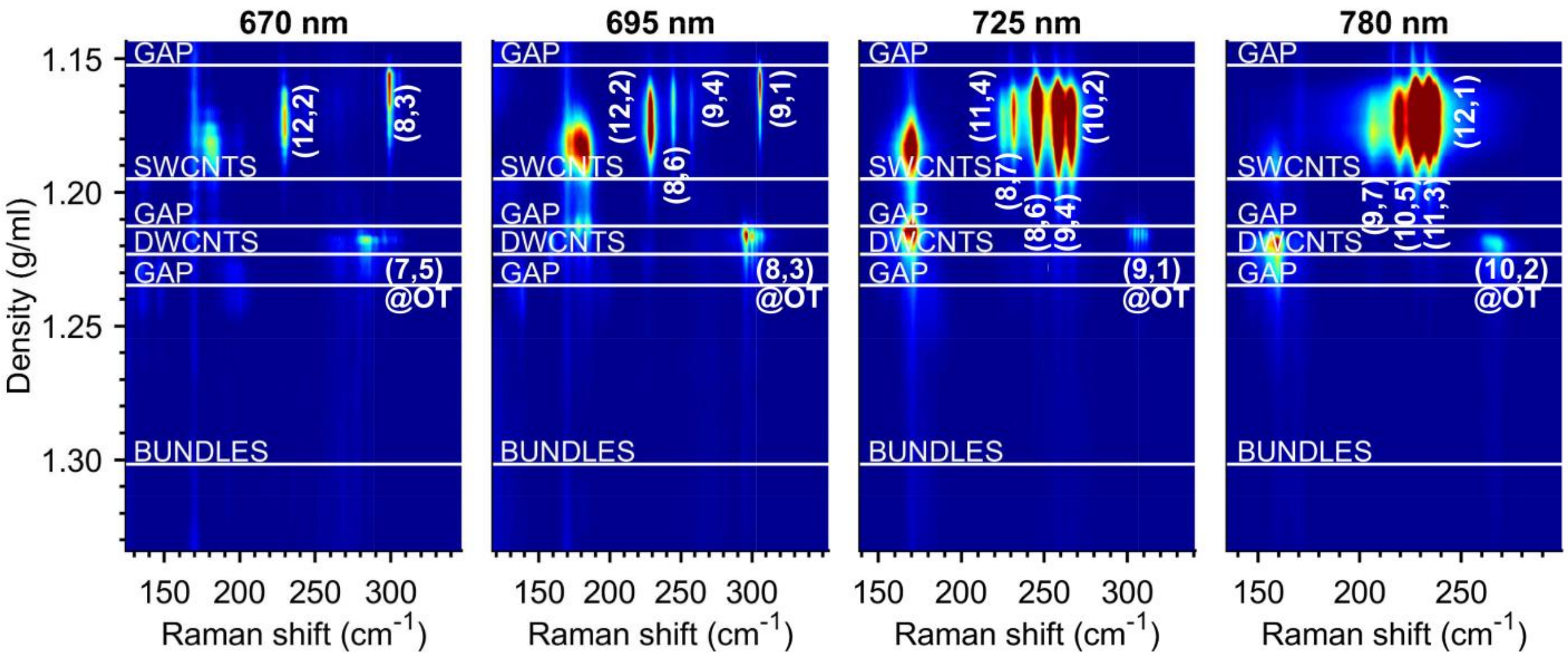

**Figure S3.** Overview of the detailed *in situ* RRS characterization as a function of density of the first DGU DWCNT purification run, measured at four different excitation wavelengths, from left to right, 670, 695, 725 and 780 nm. The color maps show how the SW- and DWCNT distributions change depending on the varying excitation wavelength, hence chiralities probed. The assigned chiralities are overlaid in white. The afterwards collected fractions, indicated by the horizontal white lines, are chosen to best fit the CNT distributions for all excitations, minimizing any overlapping regions.

## Two-step purification

Finally, the collected DWCNT fraction is similarly DGU sorted a second time to further increase the purity of the DWCNTs. To assess how much more the DWCNTs are purified from the SWCNTs by this second DGU run, the DWCNT fraction taken from DGU run 1 is compared with the DWCNT fraction from DGU run 2 through RRS measured at 725 nm. Figure S5 reveals the further decrease of the SWCNT RBMs with respect to those of the DWCNTs.

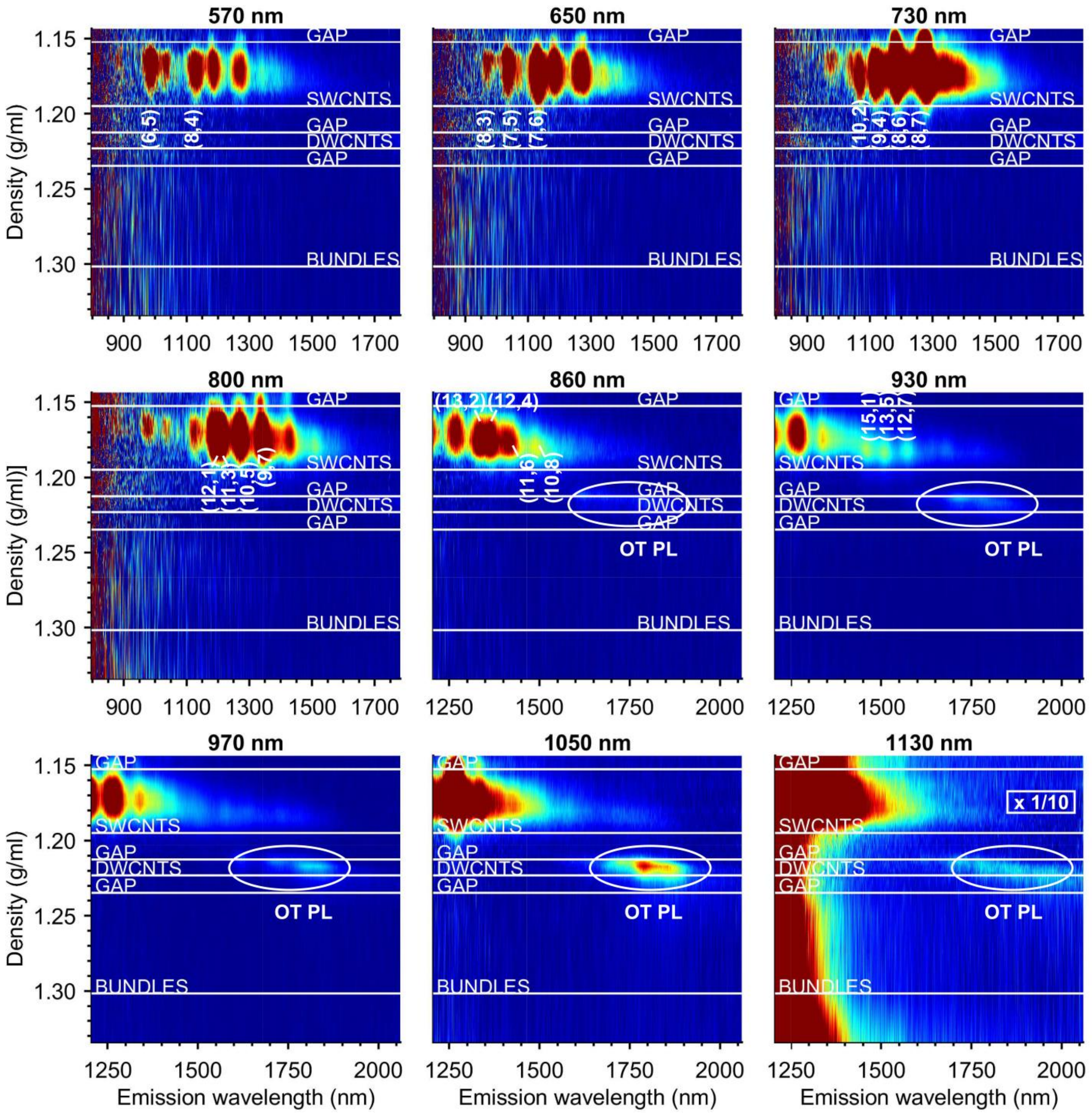

**Figure S4.** Overview of the *in situ* PL characterization as a function of density of the first DGU DWCNT purification run, measured at nine different excitation wavelengths, from left to right and top to bottom, 570, 650, 730, 800, 860, 930, 970, 1050 and 1130 nm. All color maps are shown on the same color scale, except for the last one measured at 1130 nm, which is divided by a factor of ten. In case of the five first color maps, the emission spectrometer is centered around the emission of the smaller diameter SWCNTs. The remaining four excitation wavelengths are mainly used to probe the emission of the outer tubes of the DWCNTs, but show as well the SWCNT PL via excitation in the tail of the $E_{11}$ transitions. Again, the assigned indices and the collected fractions are indicated in white on the color maps.

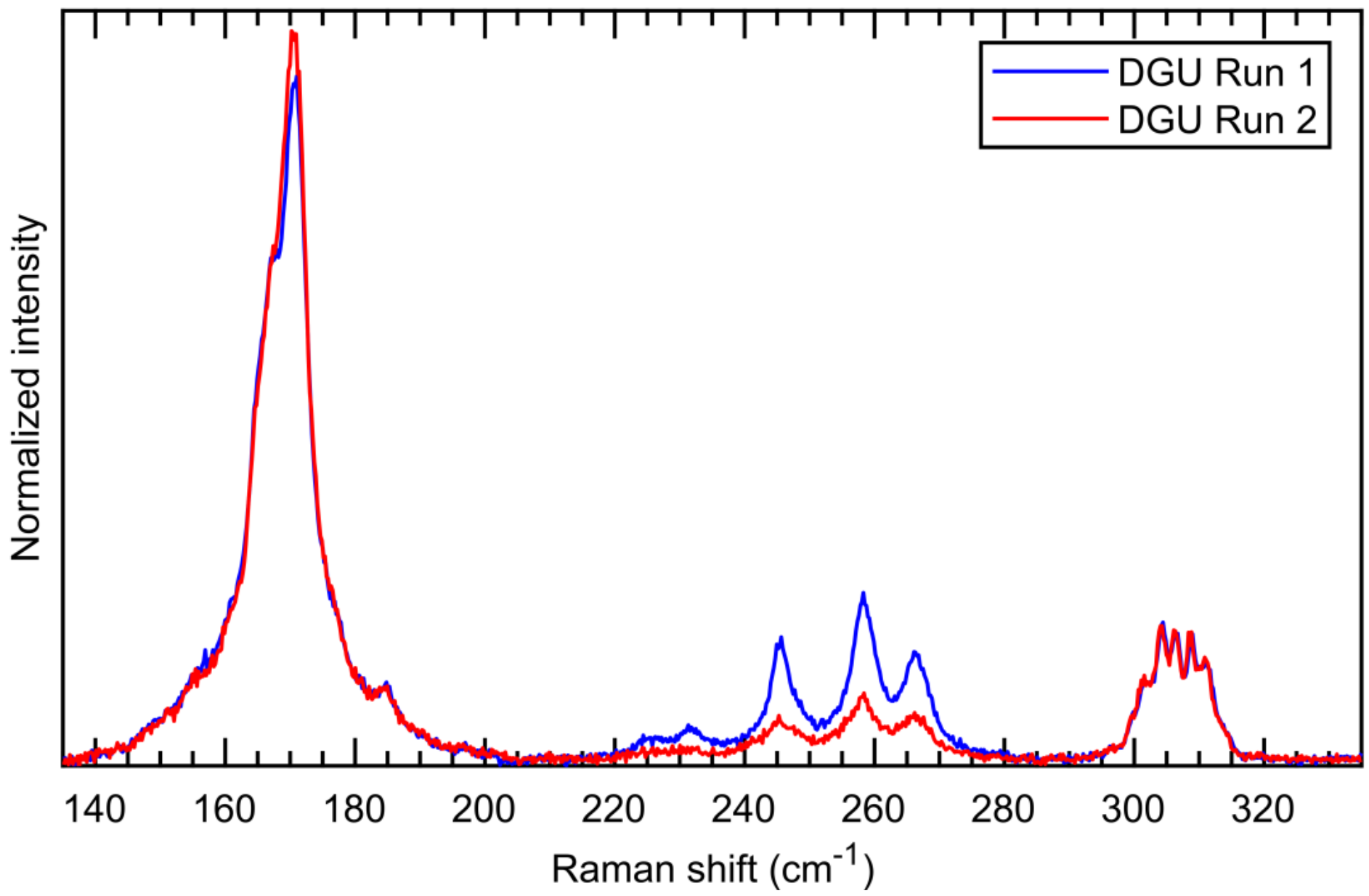

**Figure S5.** Comparison between the RRS spectra measured at 725 nm of the DWCNT fraction of DGU run 1 (blue trace) and the DWCNT fraction of DGU run 2 (red trace). After the second DGU, the DWCNTs are significantly more purified from the SWCNTs, as is visible from the decrease in SWCNT RBMs (220 – 280 cm$^{-1}$) with respect to those of the DWCNTs (140 – 200 cm$^{-1}$ and 295 – 315 cm$^{-1}$).

# Time-dependent extraction of SWCNTs by ultrasonication

## Sonication efficiency

In order to extract the inner shells from the DWCNTs, a bath ultrasonicator is used. The delivered power by such a bath ultrasonicator depends strongly on the water level and the position of the sample in the bath [2,3]. Therefore, the sonochemical efficiency of this particular device is estimated by means of KI-oxidation dosimetry, where the amount of $I_3^-$ molecules produced from KI is probed *via* absorption spectroscopy as a function of ultrasonication time [2,3]. The rate at which this $I_3^-$ concentration, $C$, increases is a measure for the sonochemical efficiency, $SE$, times the ultrasonic power, $P$, (or the relative sonochemical effective power $SE \cdot P$) over the sample's volume, $V$, following

$$\frac{C}{t} = \frac{SE \cdot P}{V}$$

In this particular experiment, 200 μL of a 0.1 M KI (Merck, ultrapure) solution was ultrasonicated in accumulating time steps and intermittently the absorption spectrum, was measured. As is demonstrated in the left panel of Figure S6, with increasing ultrasonication time distinct $I_3^-$ absorption peaks grow stronger. In order to determine $C$ from these spectra, the absorption peak at 355 nm, with a molecular extinction coefficient of $\epsilon = 26.303$ L/(mol cm) can be used [2,3]. In the right panel of Figure S6, the resulting linear ultrasonication-time-dependency of the $I_3^-$ concentration is plotted. From these results and the above relation follows that the relative sonochemical effective power per unit volume, or in other words the tri-iodide production rate, becomes $SE \cdot P/V = 9.1 \cdot 10^{-5}$ M/s which is in good agreement with typical values reported earlier for bath ultrasonication [2,3].

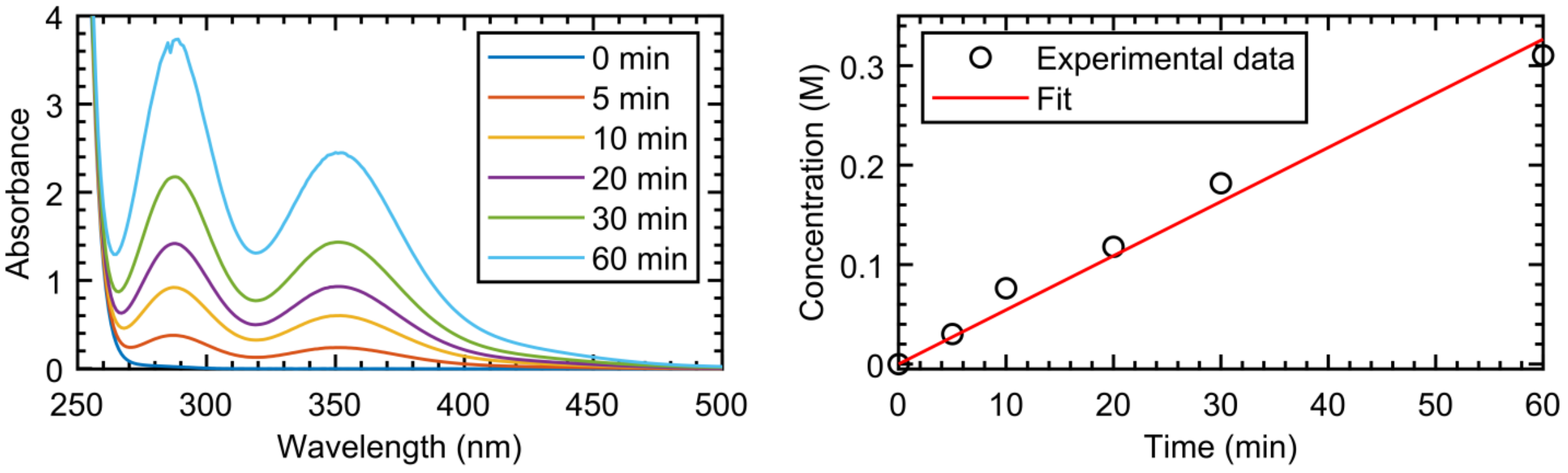

**Figure S6.** Dosimetry of the KI-oxidation reaction used to estimate the sonochemical effective power of the bath ultrasonicator used in this work. In the left panel, distinct absorption features related to the production of $I_3^-$ molecules grow stronger with increasing ultrasonication time. In the right panel, the slope of the ultrasonication-time-dependent $I_3^-$ concentration is exactly the relative sonochemical effective power per unit volume.

**Extraction rate**

To follow the ultrasonication-induced inner shell extraction mechanism as a function of ultrasonication time the purified DWCNT solution is ultrasonicated in successive time steps and intermittently characterized using RRS and PL. In case of RRS, 725 and 780 nm are used, where SW- and DWCNTs can be probed simultaneously, while for PL the SWCNTs are probed with 725 and 780 nm and the DWCNT outer shells with 970 and 1050 nm. The PL spectra (not shown in the main text) that are recorded as a function of the total ultrasonication time for excitation wavelengths 780 and 970 nm are presented in Figure S7.

To avoid any alignment-induced variations in between different experiments, we only consider the ratio of SWCNTs to DWCNTs as this ratio is unaffected by such variations. To calculate the SW- to DWCNT ratios for RRS and PL spectroscopy separately, their respective signals are integrated over fixed intervals and all contributions from different excitation wavelengths are averaged.

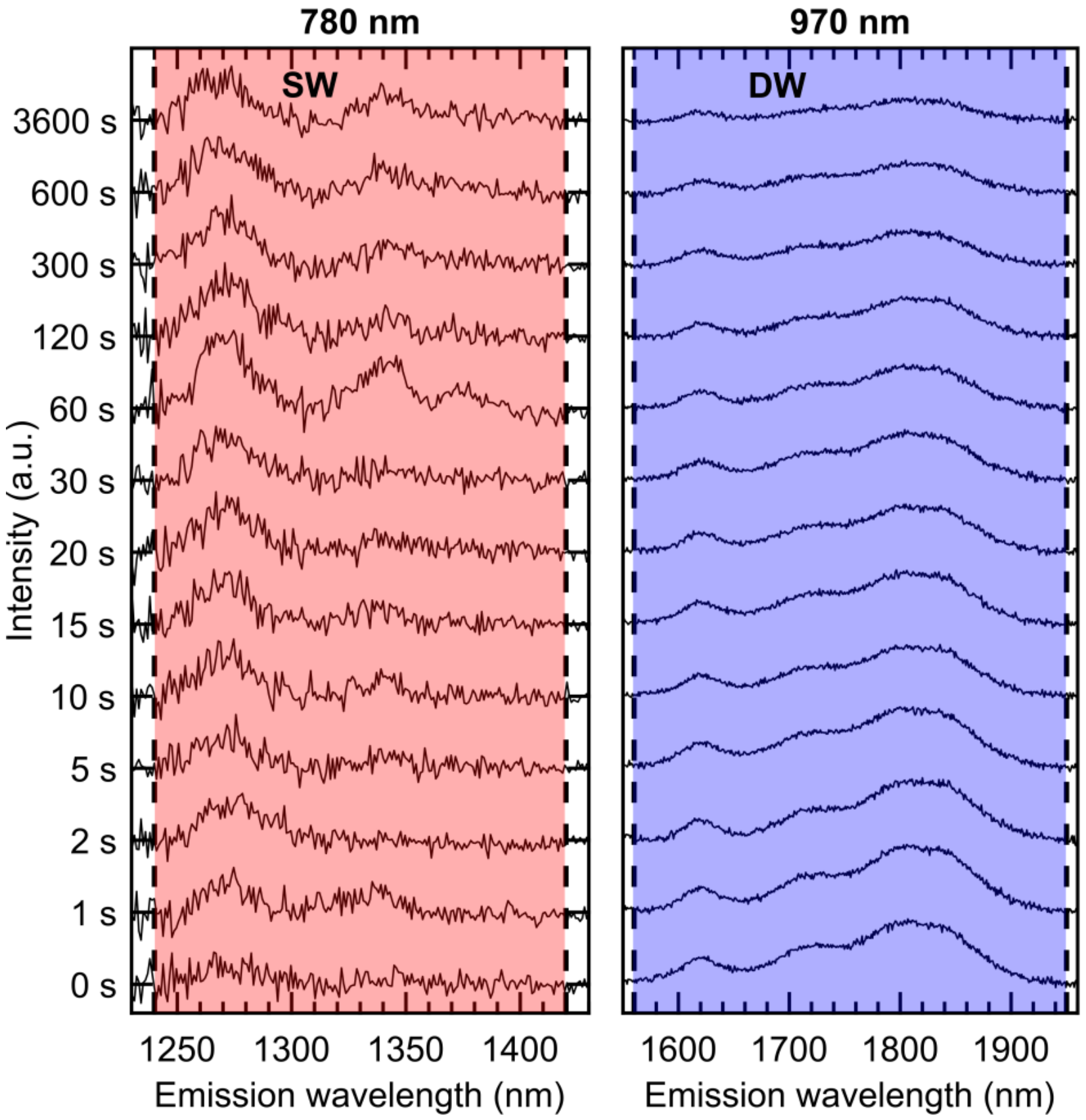

**Figure S7.** PL spectra for increasing ultrasonication time at excitation wavelengths 780 and 970 nm to probe respectively the extracted SWCNTs (red highlight) and the DWCNT outer shells (blue highlight). We observe an increase in SWCNT intensity and a decrease of the DWCNT fluorescence for accumulating ultrasonication time. Similarly to the RRS and PL spectra in the main text, at some ultrasonication times the overall intensity of the PL spectra changes suddenly, probably due to a change in the sample's alignment. The vertical dashed lines as well as the colored backgrounds indicate the integration ranges used to calculate the SW- and DWCNT intensity, respectively denoted by SW (red) and DW (blue).

For the RRS measurements, the ratio is calculated based upon the RBMs of the extracted SWCNTs, centered around 260 and 230 $cm^{-1}$ for 725 and 780 nm respectively, and the DWCNT inner shell RBMs, at 310 and 260 $cm^{-1}$ for 725 and 780 nm respectively. The individual ratios obtained for these excitation wavelengths, as well as the average of the two (black), are shown in the left most panel of Figure S8. For both excitation wavelengths a similar stretched exponentially increasing trend is observed, which can be modeled using the following fit function $r(t) = f - b\,e^{-(t/\tau)^s}$, where $f$ - $b$ and $f$ are respectively the beginning and final ratios, $\tau$ is the extraction time constant in seconds, and $s$ is the stretching parameter. The DWCNT outer shell intensities are deliberately not used in the above calculations since they are possibly complicated by overlapping outer shell SWCNT contributions. The ratio between DWCNT inner and outer shell RBMs is, however, useful to address the extraction of these otherwise inaccessible outer shell SWCNTs. The middle panel of Figure S8 shows a stretched exponentially decreasing trend for this ratio for both excitation wavelengths, meaning that the DWCNT inner and outer shell RBMs do not diminish equally fast upon ultrasonication. This effect is assigned to an increase of the SWCNT outer shell RBMs, which overlap with the DWCNT outer shell Raman modes. Hence, the ultrasonication-induced extraction process forms not only inner but also outer shell SWCNTs. Finally, a particularly interesting comparison is made in the right panel of Figure S8 where the extracted inner shell SWCNT intensity of one specific chirality is compared with the RBM intensity of that exact same chirality as a DWCNT inner shell. Due to the strong electronic shift of the Raman mode resonance upon encapsulation by an outer tube, this is possible for the (10,2) CNT. This chirality is in resonance with 725 nm as an individual SWCNT, while at 780 nm the (10,2) DWCNT inner shell RBMs can be measured, as has already been indexed in Figure S3. Although this ratio is much noisier because it is obtained from two different excitation wavelengths, it increases similarly to the total SW-DWCNT ratios.

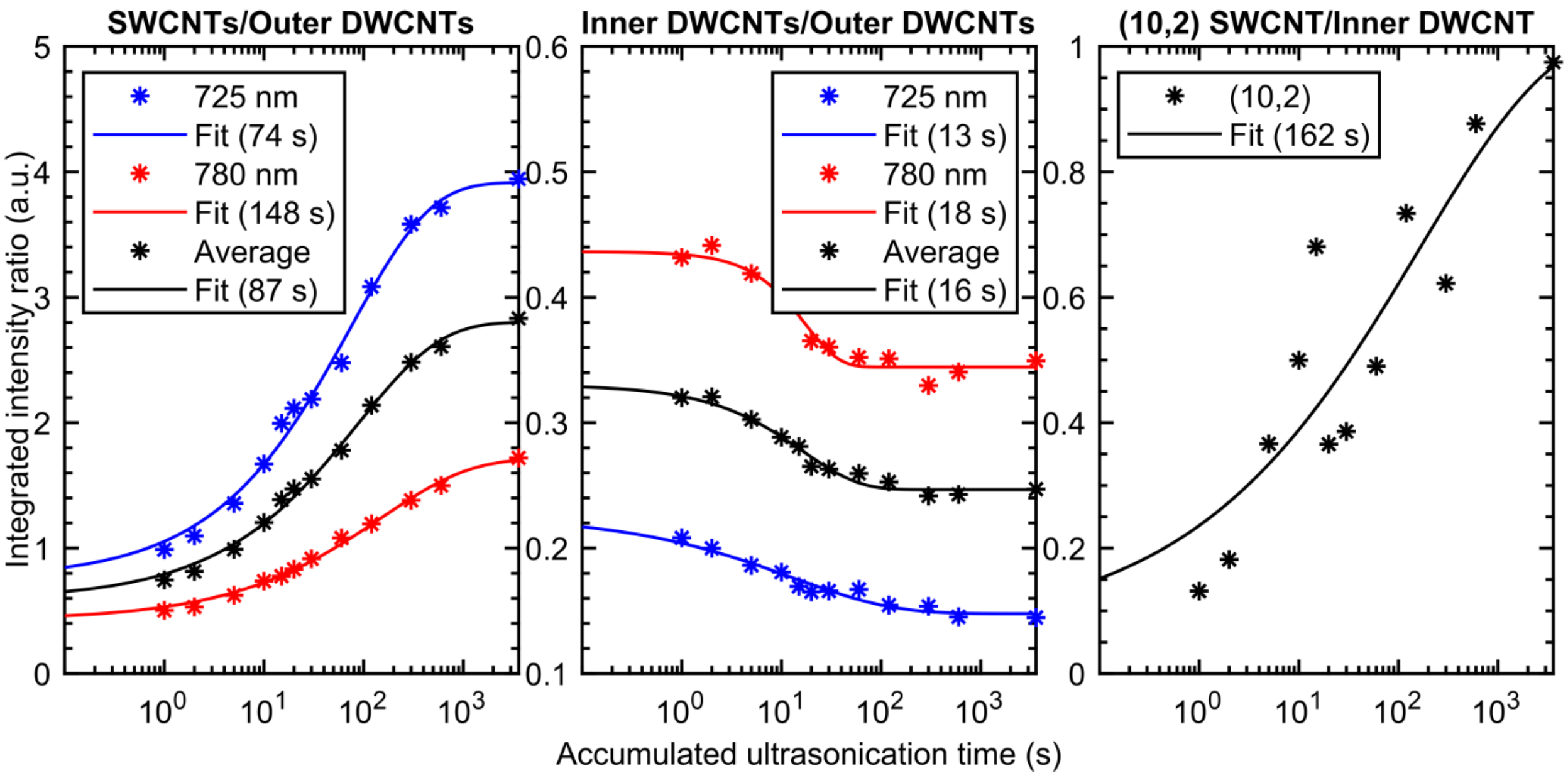

**Figure S8.** Detailed comparison of different signal ratios as a function of ultrasonication time as measured by RRS for the two individual excitation wavelengths 725 nm (blue) and 780 nm (red), as well as their average (black). The left panel shows the ratio between the extracted SWCNTs and the DWCNT inner shells. The middle panel plots the ratio of the DWCNT inner shells with respect to the outer shells. The right panel shows the intensity ratio of the (10,2) chirality between the extracted SWCNT, measured at 725 nm, and the DWCNT inner shell, measured at 780 nm (see Figure S3). All ratios are plotted on a logarithmic timescale to visualize the stretched exponential behavior, which is used as the model to fit these data.

In case of the PL spectra, the ratio is defined by the inner shell SWCNT fluorescence, measured with 725 and 780 nm, and the DWCNT outer shell fluorescence, measured with 970 and 1050 nm. The integration ranges used for both techniques and for all excitation wavelengths are indicated using vertical dashed lines and the colored backgrounds.

Figure S9 compares the resulting SW- to DWCNT ratios obtained by PL spectroscopy as a function of ultrasonication time with the ratio obtained earlier by RRS (black curve in the left most panel of Figure S8). Analogously to the RRS ratio, the ratio obtained by PL spectroscopy follows a stretched exponentially increasing model which can be fitted similarly. Table S1 summarizes all of the obtained fit parameters for both ultrasonication time-dependent ratios. From this comparison follows that the PL ratio increases more slowly than the ratio obtained from RRS, although both time constants have a similar order of magnitude, which could be attributed to a difference between the Raman cross sections and PL quantum efficiencies of the tubes probed or possibly to a length-dependent effect on the PL quantum efficiencies.

To estimate how many of the DWCNTs are being pulled apart and, hence, are destroyed during the extensive ultrasonication process we compare the intensities of the DWCNT signals before and after ultrasonication in both RRS and PL at several excitation wavelengths. Similarly to the ultrasonication time-dependent intensity ratios, the DWCNT signal intensity is calculated by integrating the spectra before and after ultrasonication over a fixed range. The results summarized in Figure S10 show that after ultrasonication the total intensity of the DWCNT inner shell RBMs is decreased to 40% of its initial value, whereas the PL intensity is decreased to 30% compared to the starting PL intensity. Hence, although the extraction rate in PL appears to be slower than in RRS, the total effect of the ultrasonication-induced extraction is more severe for the DWCNT fluorescence than it is for the RBM intensity. This difference between the total amount of destroyed DWCNTs in RRS and PL could also be explained by the difference between the Raman cross sections and the PL quantum efficiencies of the tubes in their SW- or DWCNT configuration as well as additional quenching effects by defects and filling in both PL and RRS.

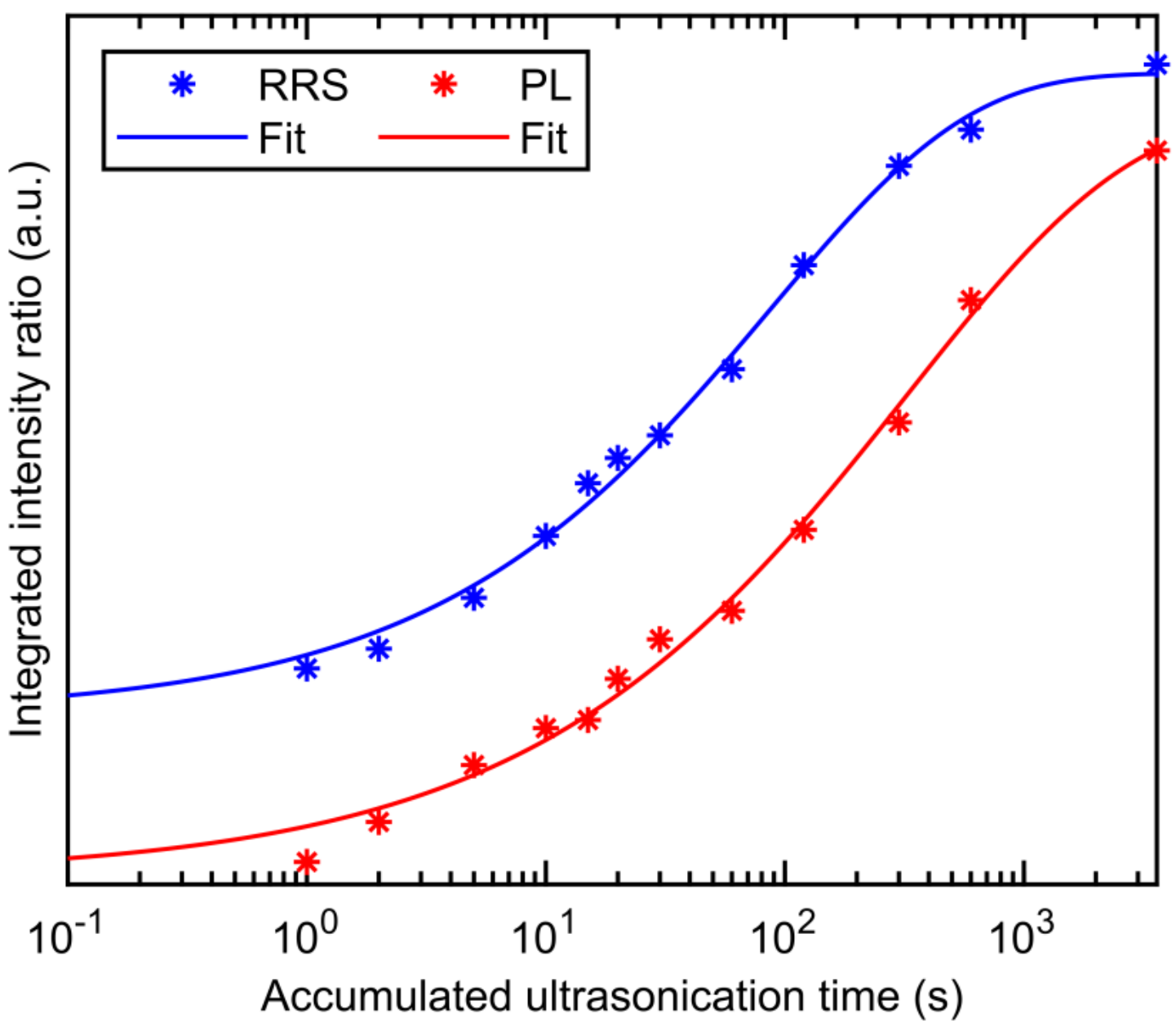

**Figure S9.** The ratios of SW- to DWCNT intensity as a function of accumulative ultrasonication time. The SW- and DWCNT intensities are calculated by integrating their respective signals over fixed intervals and averaging the contributions over all probed excitation wavelengths. The ratios are plotted on a logarithmic timescale to show the stretched exponentially increasing behavior of the SW- to DWCNT ratios measured by both RRS (blue) and PL (red).

| **Fit parameters** | Beginning ratio $f - b$ | Final ratio $f$ | Time constant (s) $\tau$ | Stretching parameter $s$ |
|---|---|---|---|---|
| RRS | 0.6± 0.2 | 2.8 ± 0.1 | 87 ± 23 | 0.52 ± 0.08 |
| PL | 0.03 ± 0.4 | 2.7 ± 0.3 | 327 ± 164 | 0.47 ± 0.09 |

**Table S1.** Overview of the obtained parameters for the exponentially stretched fits of the SW- to DWCNT ratios in case of RRS and PL spectroscopy shown in Figure S8.

Finally, as ultrasonication is known to damage CNTs, more defects are expected in the sample after the ultrasonication-induced extraction process. To roughly estimate how severely the sample is defected during the extraction, the intensity of the D-band is monitored relative to the intensity of the G-band before and after ultrasonication. Probing this so-called D/G-ratio at an excitation wavelength of 725 nm reveals an increase from 0.13 to 0.21 when using the integrated intensity values, while the peak-to-peak intensity ratio also increases from 0.08 to 0.10, see Figure S11. Hence, the CNTs in resonance at this particular wavelength are indeed more defective after the extraction.

The effect of ultrasonication on the DWCNTs can also be seen with absorption spectroscopy. However, due to the large scattering background, this is not so obvious. Therefore, the spectra shown in Figure 7 of the main text are first background-corrected by fitting the π-plasmon band around 270 nm with a Lorentzian and the Mie-scattering background with an inverse power law, $\boldsymbol{a}/\boldsymbol{\lambda}$, where $\boldsymbol{a}$ is a fitting parameter and $\boldsymbol{\lambda}$ the absorption wavelength. The resulting fitted backgrounds of the non-sonicated purified DWCNTs and the ultrasonicated DWCNTs are shown in Figure S12.

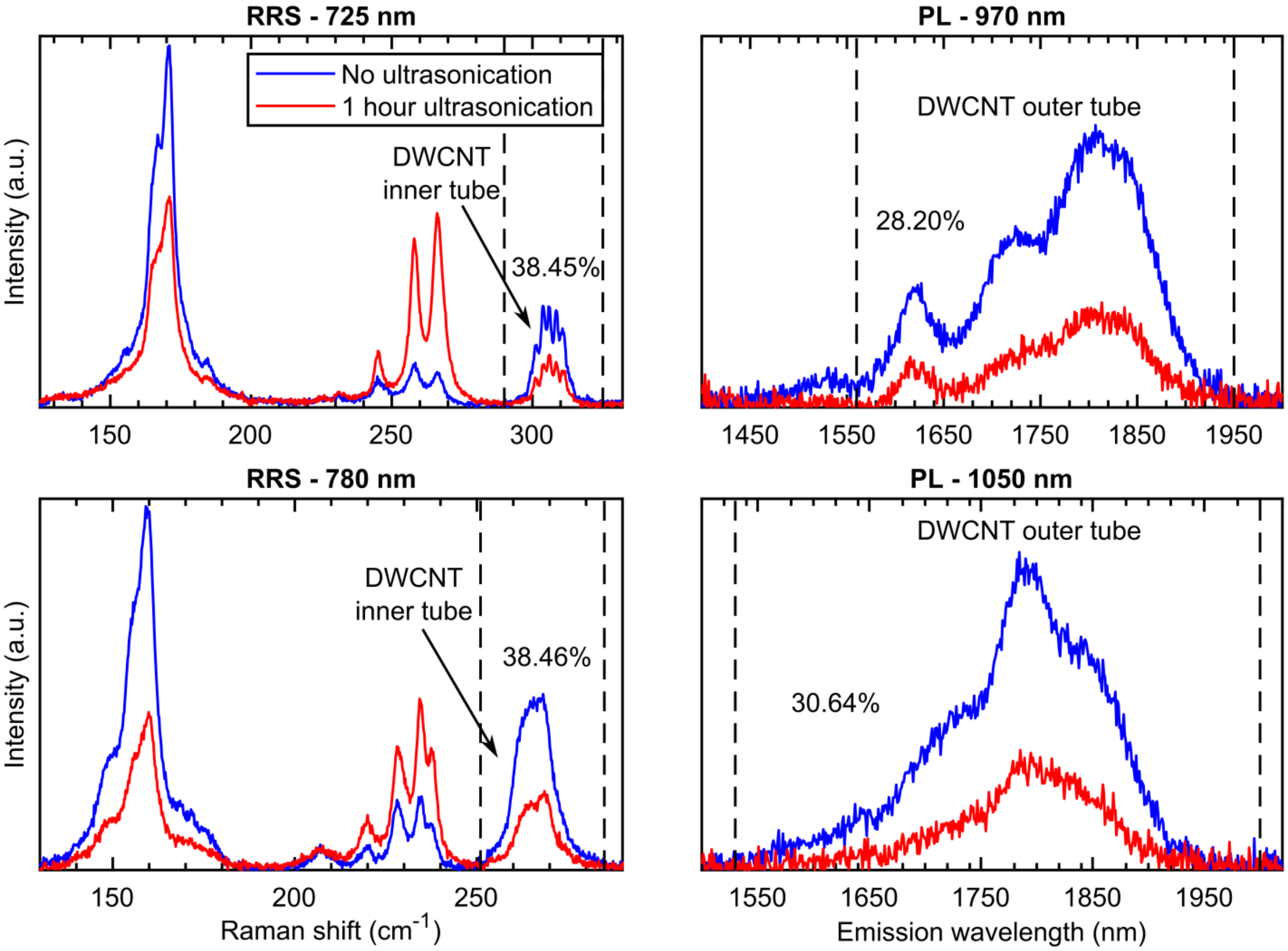

**Figure S10.** Based upon the integrated DWCNT signal intensities before and after ultrasonication, the DWCNT inner shell RBM intensity is decreased to 40% of the initial intensity after ultrasonication (left column), while the DWCNT outer shell PL is only 30% of that of the starting intensity (right column). For RRS, excitation wavelengths 725 (top left) and 780 nm (bottom left) and the DWCNT inner shell RBM signals are used. For PL, excitation wavelengths 970 (top right) and 1050 nm (bottom right) and the DWCNT outer shell PL signals are used. The vertical dashed lines indicate the integration ranges.

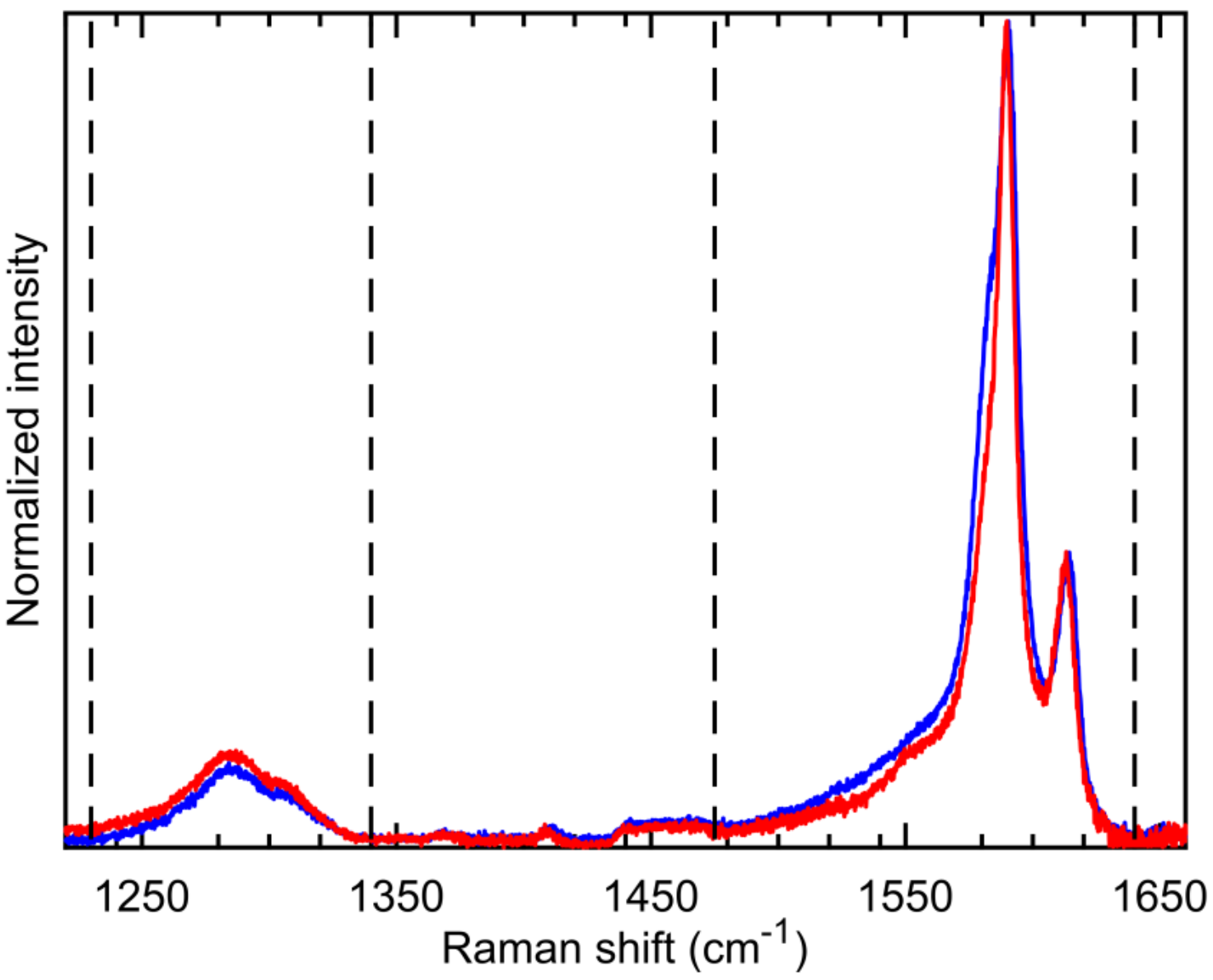

**Figure S11.** RRS spectra of the D- and G-band, normalized on the G-band's peak intensity, measured at 725 nm, before (blue trace) and after (red trace) the extensive ultrasonication-induced extraction process. The intensities of both bands are determined by integrating the signals over the range indicated by the vertical dashed lines, resulting in an increasing D/G-ratio from 0.13 to 0.21. The peak-to-peak D/G-ratio also increases, from 0.08 to 0.10. Hence this reveals that the CNTs in resonance are indeed more defective after ultrasonication.

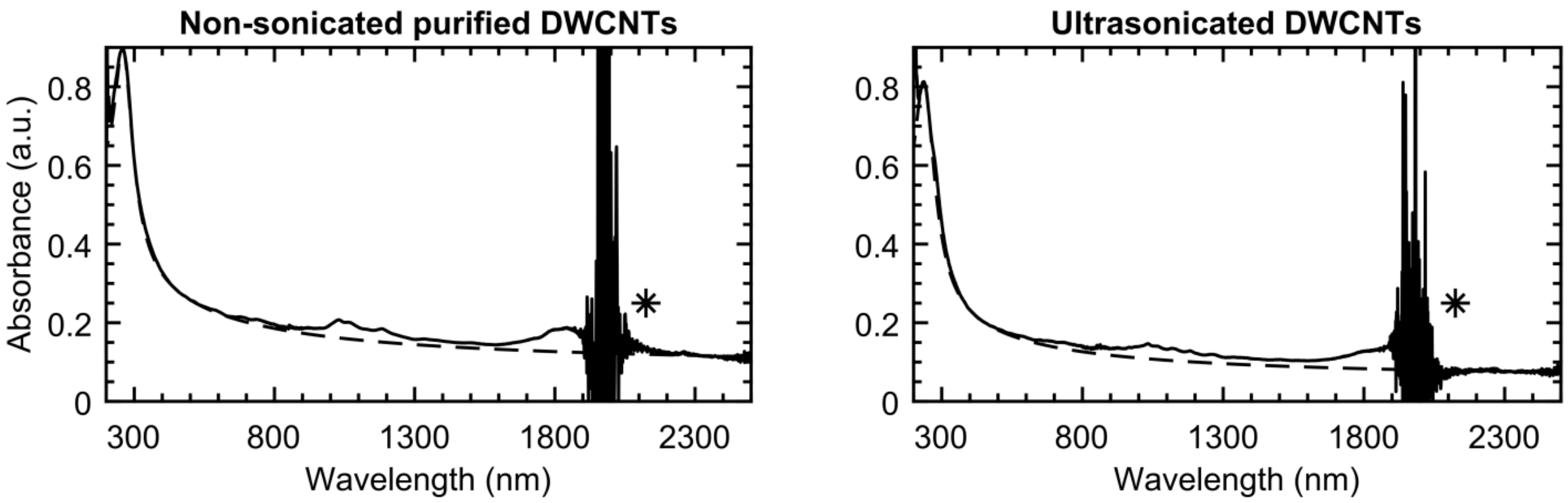

**Figure S12.** Comparison of the raw absorption spectra (solid lines) and the fitted backgrounds (dashed lines) in case of the purified DWCNTs before (left) and after (right) extensive ultrasonication. The asterisk denotes the noisy region that originates from the saturated $D_2O$ baseline correction.

## DGU sorting of the extracted SWCNTs

Once the inner shell SWCNT extraction is saturated (one hour), the extracted SWCNTs are separated from the remaining DWCNTs using DGU to prove that these SWCNTs are indeed fully extracted from the DWCNTs and individually solubilized. The separation is done similarly to the previous DGU sorting and is characterized again by *in situ* RRS and PL spectroscopy at various excitation wavelengths. Together with the ultrasonicated DWCNTs, the unaffected purified DWCNT solution is DGU sorted as well to act as a reference.

Analogously to the first DGU run the density-to-depth relation in the ultracentrifuge tube is determined. This time there are two spare ultracentrifuge tubes available in the DGU rotor that can be used for this purpose. The averaged density-to-depth relation is depicted in Figure S13, where the depth range probed in the following *in situ* measurements is indicated by the dashed lines.

The *in situ* RRS and PL characterization of the sorted DWCNT samples are performed at multiple excitation wavelengths to probe several sets of CNT chiralities. Based upon the excitation wavelengths used during the first DGU run, we excite at 695, 725 and 780 nm for RRS, and at 650, 970 and 1050 nm for PL. For both RRS and PL, the data of the unaffected purified DWCNTs are compared with the data of the ultrasonicated DWCNTs. Figure S14 and Figure S15 summarize these *in situ* RRS and PL measurements as a function of density, respectively. On these RRS and PL color maps, we can see in both samples the typical DWCNT features at high density (deeper in the ultracentrifuge tube). For RRS, this means the DWCNT outer and inner shell RBMs at lower and higher Raman shifts, respectively, while in case of PL only fluorescence from the DWCNT outer shells is detected at excitation wavelengths 970 and 1050 nm. Furthermore, the buoyant densities of the DWCNTs in the ultrasonicated sample match well with the unaffected DWCNTs in the reference sample.

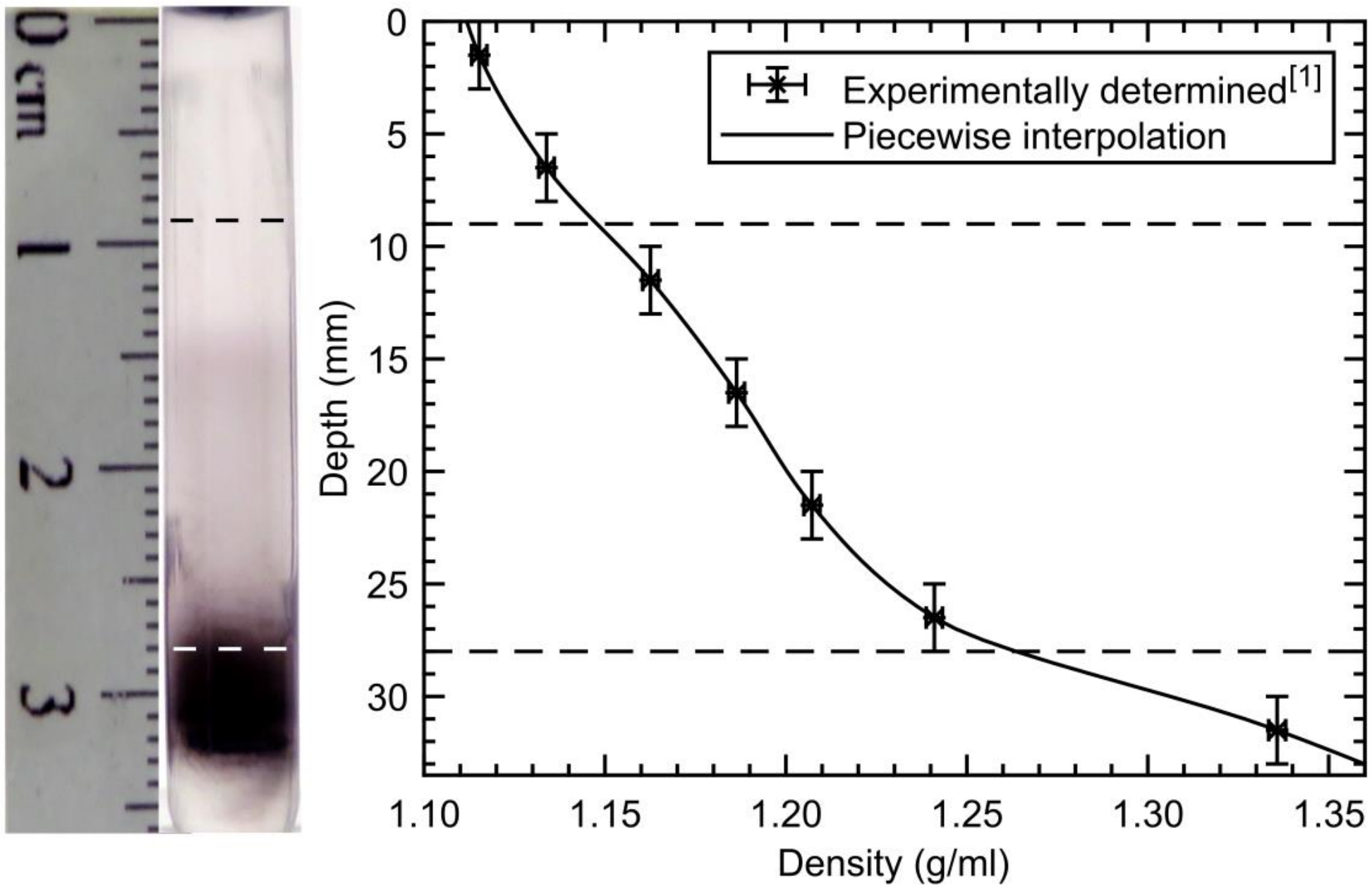

**Figure S13.** Density-to-depth relation inside the ultracentrifuge tubes after the CNTs reach isopycnic equilibrium with the gradient medium in DGU run 3, determined by examining two ultracentrifuge tubes as described in reference [1]. The data points are interpolated with a piecewise cubic Hermite polynomial. The horizontal dashed lines indicate the *in situ* probed depth range, in which the CNT distributions of interest lie in the more linear regime (SWCNTs from 10 to 19 mm and the DWCNTs from 22 to27 mm).

In case of the ultrasonicated DWCNTs, additional SWCNT RRS and PL signals are detected at a significantly lower density that matches well with the density of the as-synthesized SWCNTs displayed in Figure S3 and Figure S4. These SWCNT signals are exactly the ultrasonication-induced extracted SWCNTs. Since the extracted SWCNTs have a significantly lower buoyant density than the DWCNTs and there are no signals in between the SW- and DWCNTs, all of the extracted SWCNTs are individually solubilized and no partially extracted SW-DWCNT structures exist.

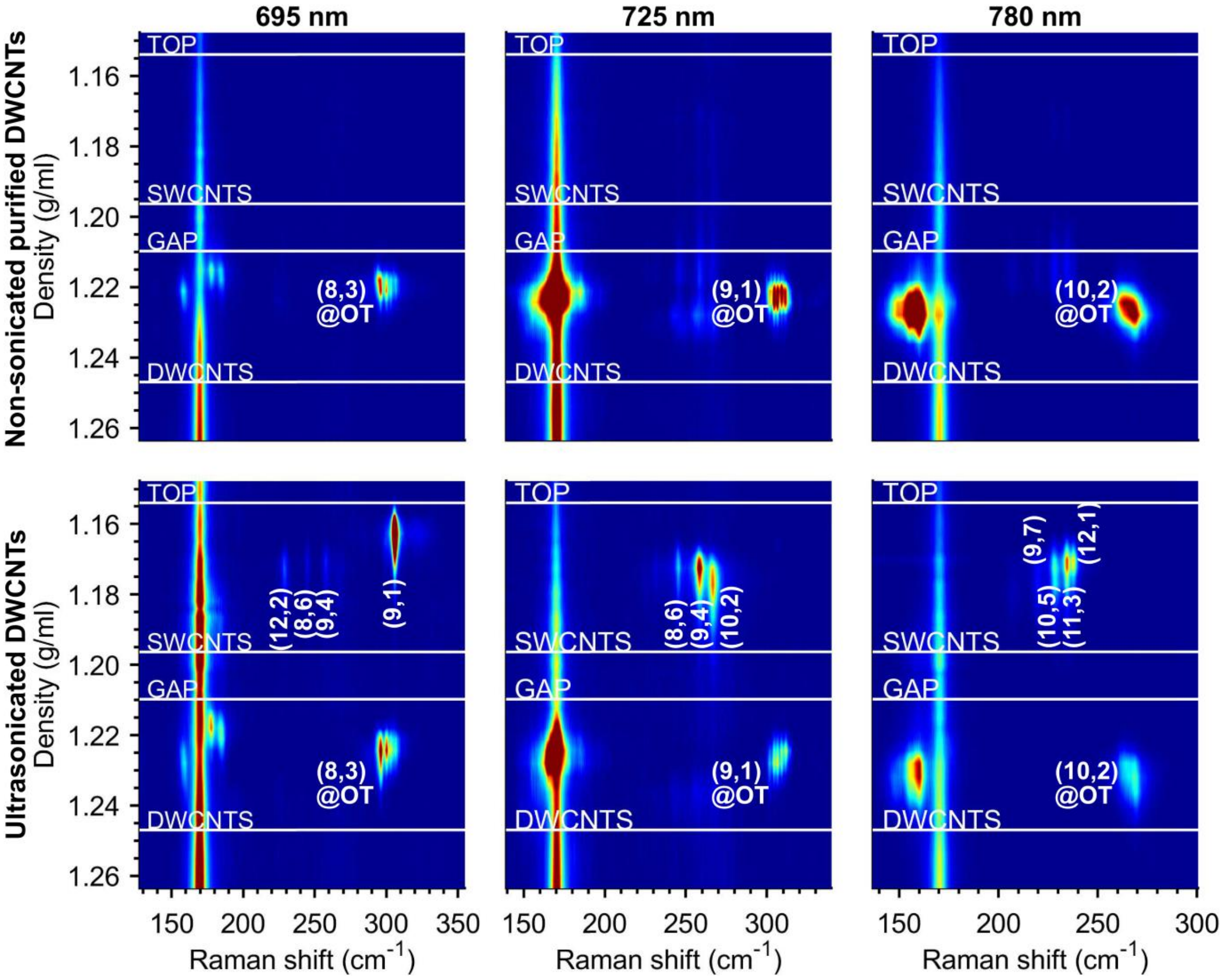

**Figure S14.** *In situ* RRS characterization of the DGU sorting of the purified DWCNTs (top panels) and the ultrasonicated DWCNTs (bottom panels) for the excitation wavelengths from left to right 695, 725 and 780 nm. In both samples the DWCNT features can be recognized at higher densities, while only for the ultrasonicated sample SWCNTs are present in the color maps at lower densities. The densities at which the fractions are collected afterwards as well as the assigned chiralities are indicated in white.

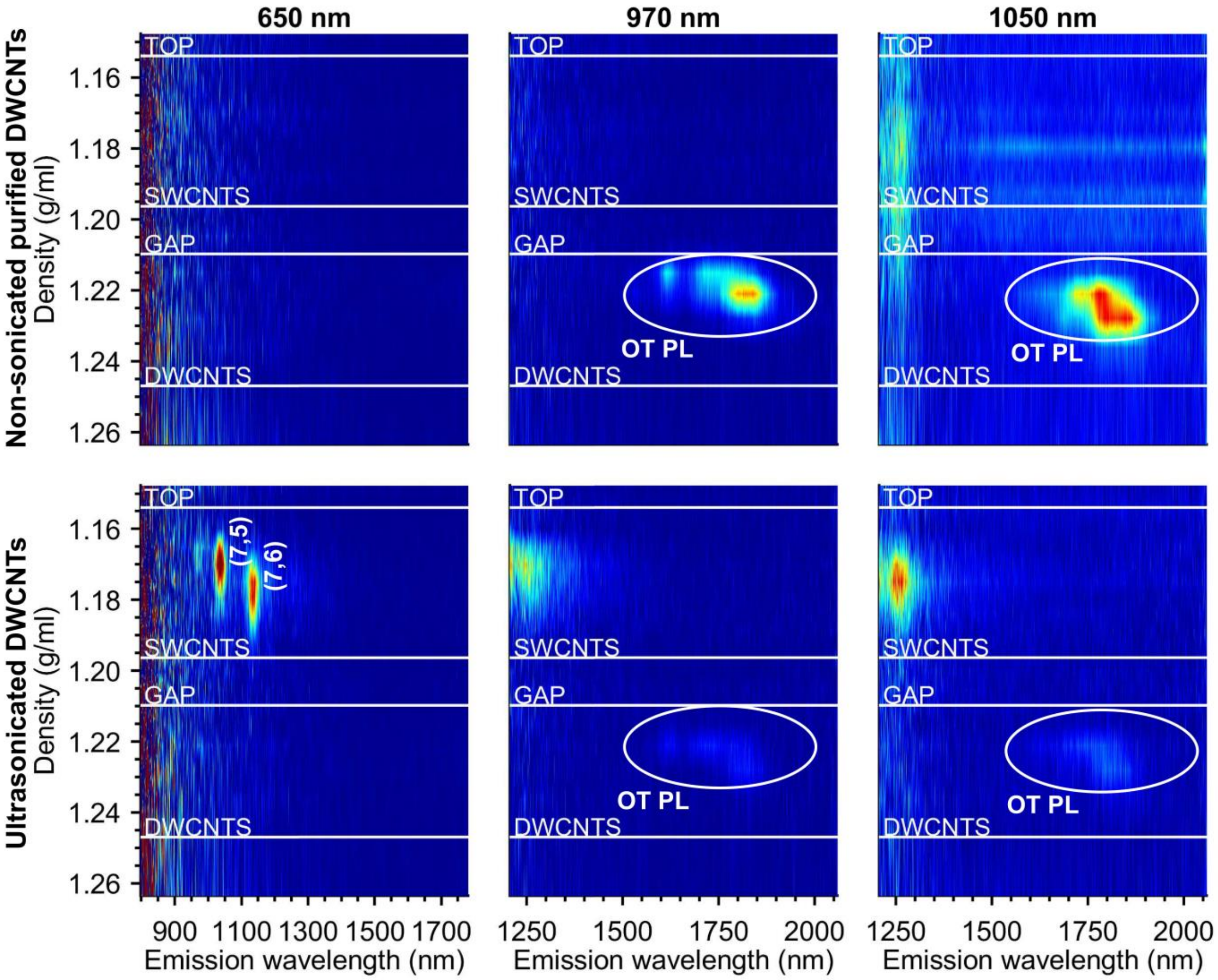

**Figure S15.** Overview of the in situ PL characterization of the DGU sorted purified DWCNTs (top panels) and ultrasonicated DWCNTs (bottom panels) measured at 650, 970 and 1050 nm from left to right. The purified DWCNTs only show PL signals at higher densities originating from the DWCNT outer shells, while in case of the ultrasonicated sample these signals are almost completely absent. Furthermore, due to the ultrasonication strong fluorescence is measured at lower densities coming from the extracted inner shell SWCNTs. The assigned fractions at which the samples are separated later on are indicated in white on the color maps.

## Transmission electron microscopy

The extracted SWCNTs in the ultrasonicated DWCNT sample are expected to be drastically shortened due to the extensive ultrasonication. Standard supporting TEM grids use holey carbon overlaid on a Cu mesh grid on which the CNTs are deposited. However, the rinsing that is necessary to remove the otherwise obscuring surfactant molecules causes especially these shortened CNTs to be flushed away through the pores of the supporting grid. This not only dramatically limits the number of observations to be taken, but also affects the observed diameter distributions as the smallest CNTs in which we are interested will be more easily lost than the larger CNTs. Therefore, specific supporting TEM grids were used that consist of lacey carbon overlaid by graphene on top of a Cu mesh grid on top of which the surfactant solubilized CNTs are deposited, as is schematically illustrated in Figure S16.

Compared to normal porous supporting grids, these graphene-covered grids efficiently keep the CNTs on the grid during the rinsing. Figure S17 illustrates this effect by comparing the statistical HRTEM data of the purified and ultrasonicated DWCNTs for the regular supporting grids with holey carbon in the left two columns and the lacey carbon grids overlaid with graphene in the right two columns. Unlike in the main text, no diameter restrictions are applied here and the abundances shown are thus valid for the entire diameter range. From top to bottom, the panels show the abundances of the different CNT species, the diameter distribution of the DWCNT inner and outer shells and the distribution of the SWCNTs. In case of the regularly used holey carbon grids, the amount of observations, *N*, in case of the ultrasonicated DWCNT sample is only 52, which is too little to draw any statistically meaningful conclusions. The graphene covered TEM grids, on the contrary, have well over 300 CNT observations, allowing for a far more accurate statistical result. Even in case of the purified DWCNT sample, more observations are easily made with these grids. Moreover, a lot more SWCNTs with a diameter below 1 nm are detected when using the graphene coated grids, hence proving that they effectively keep even the smallest CNTs on the grids.

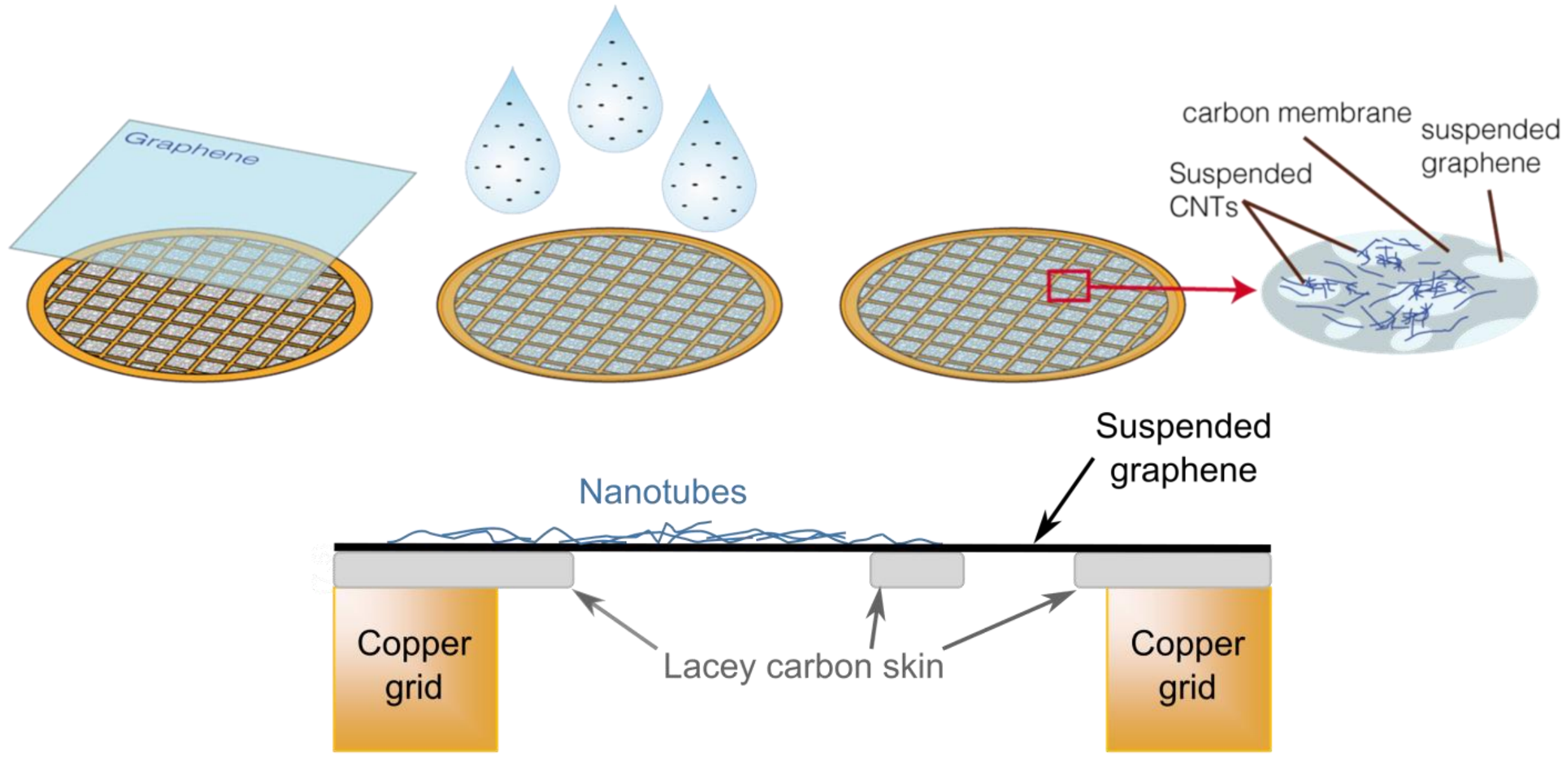

**Figure S16.** Schematic representation of how the surfactant solubilized CNTs are deposited on top of a supporting grid for HRTEM observations consisting of graphene covered lacey carbon on top of Cu. The graphene covers the holes in the lacey carbon, preventing hereby the smallest CNTs from being flushed away by the rinsing during the sample preparation.

In addition to the minimized loss of CNTs, the additional graphene layer does not significantly affect the image quality due to its atomic thickness, keeps the CNTs in focus and reduces charge effects because of its metallic behavior. Taking into account that on this type of grid SWCNTs with a diameter around 0.6 nm are observed, which was not the case when using the normal grids, it is thus a good choice of supporting grid for such delicate CNT samples as the ones studied in this work.

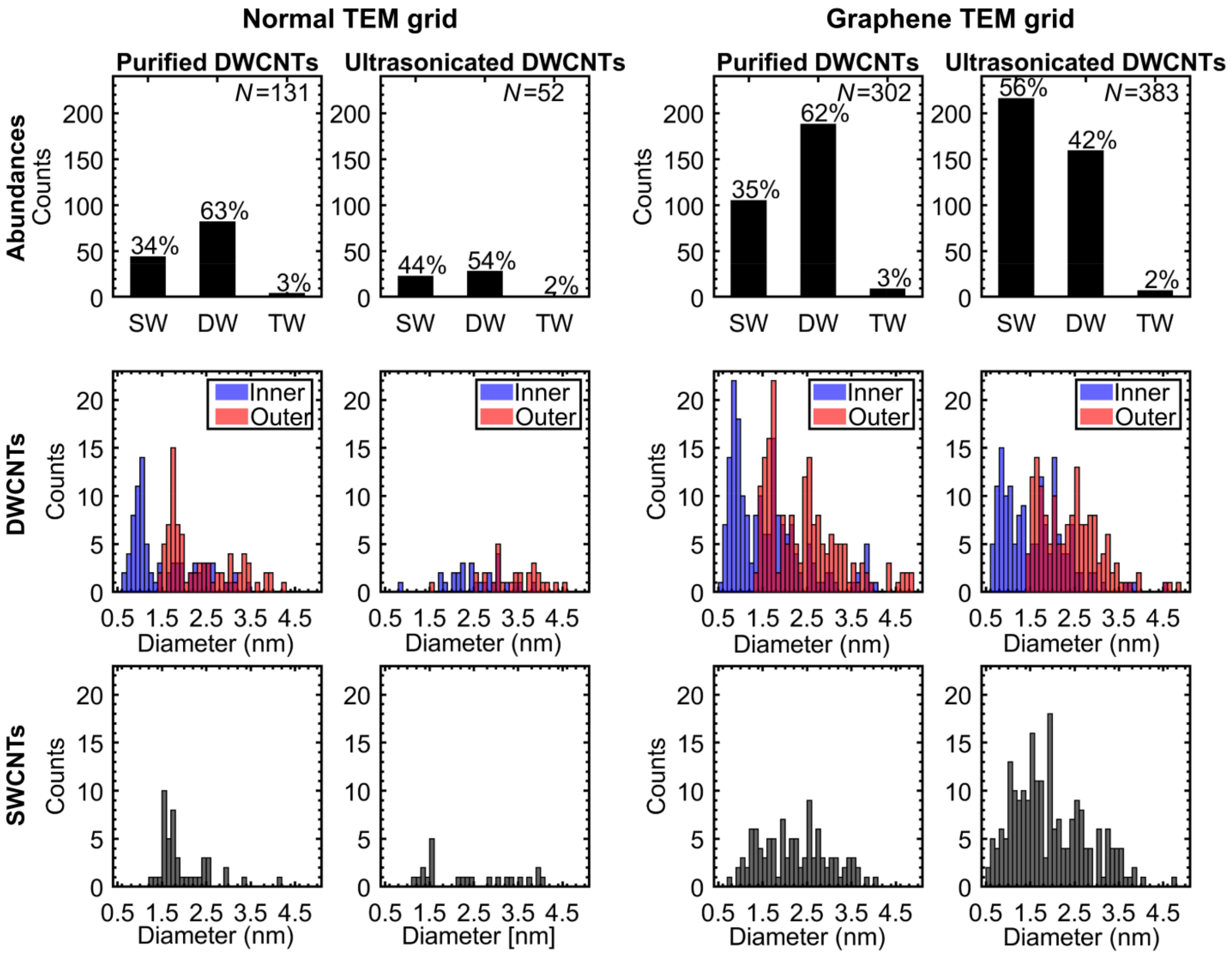

**Figure S17.** Overview of the HRTEM results of the purified and ultrasonicated DWCNT samples for the normal, holey carbon TEM grids on the left and the graphene covered TEM grids on the right. From top to bottom, the different panels show the abundances of the CNT species, the diameter distributions of the DWCNT inner and outer shells and the distribution of the SWCNTs. When the graphene-coated grids are used many more CNT observations (*N*) are made, which results in far more accurate statistical results.

According to the results in Figure S17, more SWCNTs are observed after ultrasonication in the diameter range of both the DWCNT inner as well as outer shells. Contrarily to optical spectroscopy, HRTEM is thus successful in directly observing the extracted outer shell SWCNTs. However, upon observation it became apparent that in case of the ultrasonicated DWCNT sample significantly more CNTs showed defects and were truncated compared to the more intact CNTs in the purified DWCNT sample. Figure S18 and Figure S19 present an overview of several representative HRTEM images of both samples. Figure S18 shows how drastically shortened the CNTs are after ultrasonication, hence the reason why they are so easily flushed through the holes if normal supporting grids are used. Figure S19 illustrates that many of the ultrasonicated CNTs show abnormal structures, including holes in the sidewalls, or are even cut, which is in line with the previously recorded increased D/G-ratio. Since in a DWCNT the outer shell shields the inner tube from the ultrasonication power, the outer tubes are expected to be more damaged by the ultrasonication than the inner tubes. Together with the fact that such large-diameter SWCNTs are anyway harder to detect spectroscopically, this can explain why these extracted large-diameter SWCNTs were so difficult to spectroscopically detect in this work compared to the smaller diameter inner shell SWCNTs.

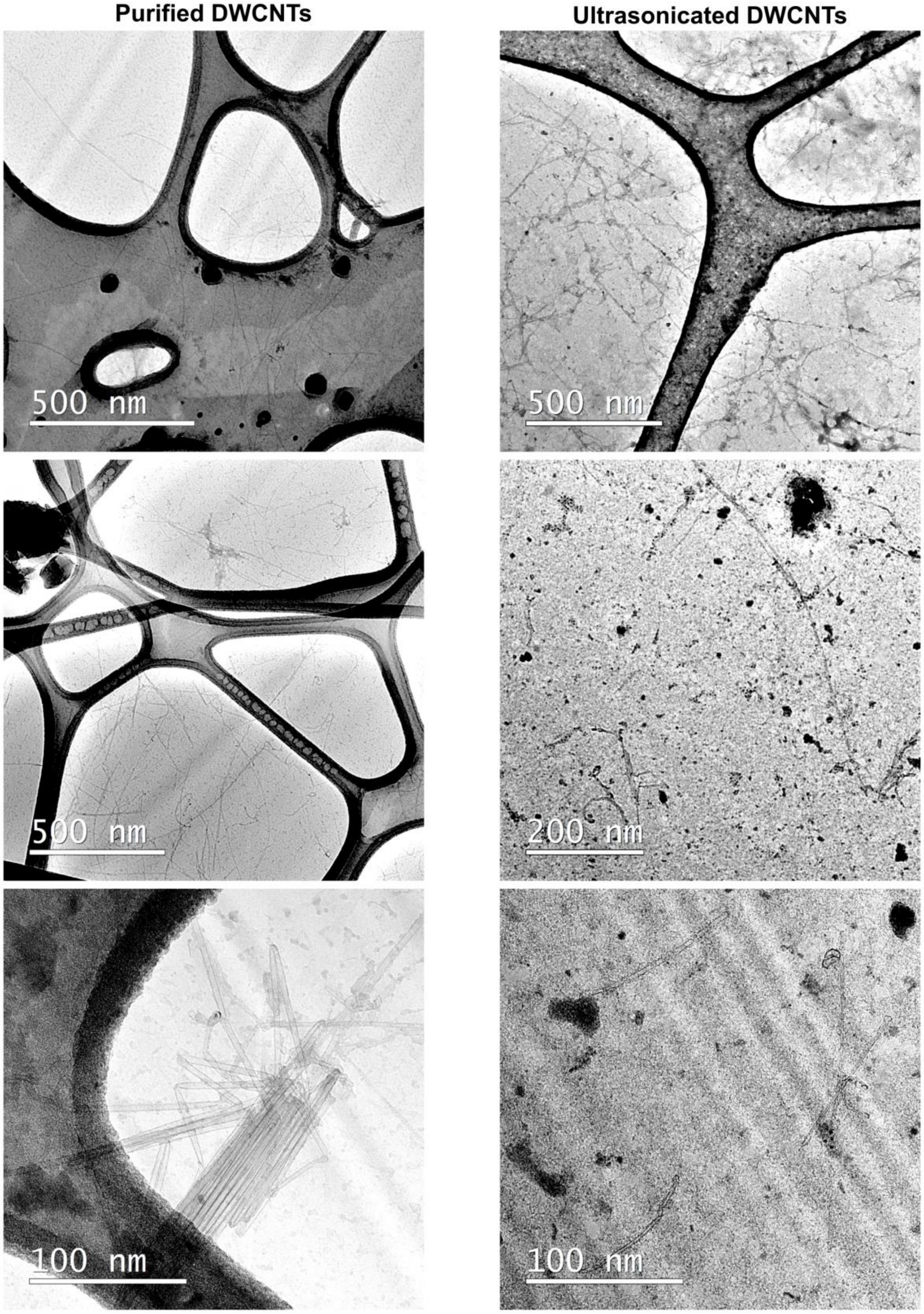

**Figure S18.** Overview of several representative HRTEM images of the purified (left) and ultrasonicated (right) DWCNT sample at low magnification, illustrating the ultrasonication-induced drastic shortening of the CNTs.

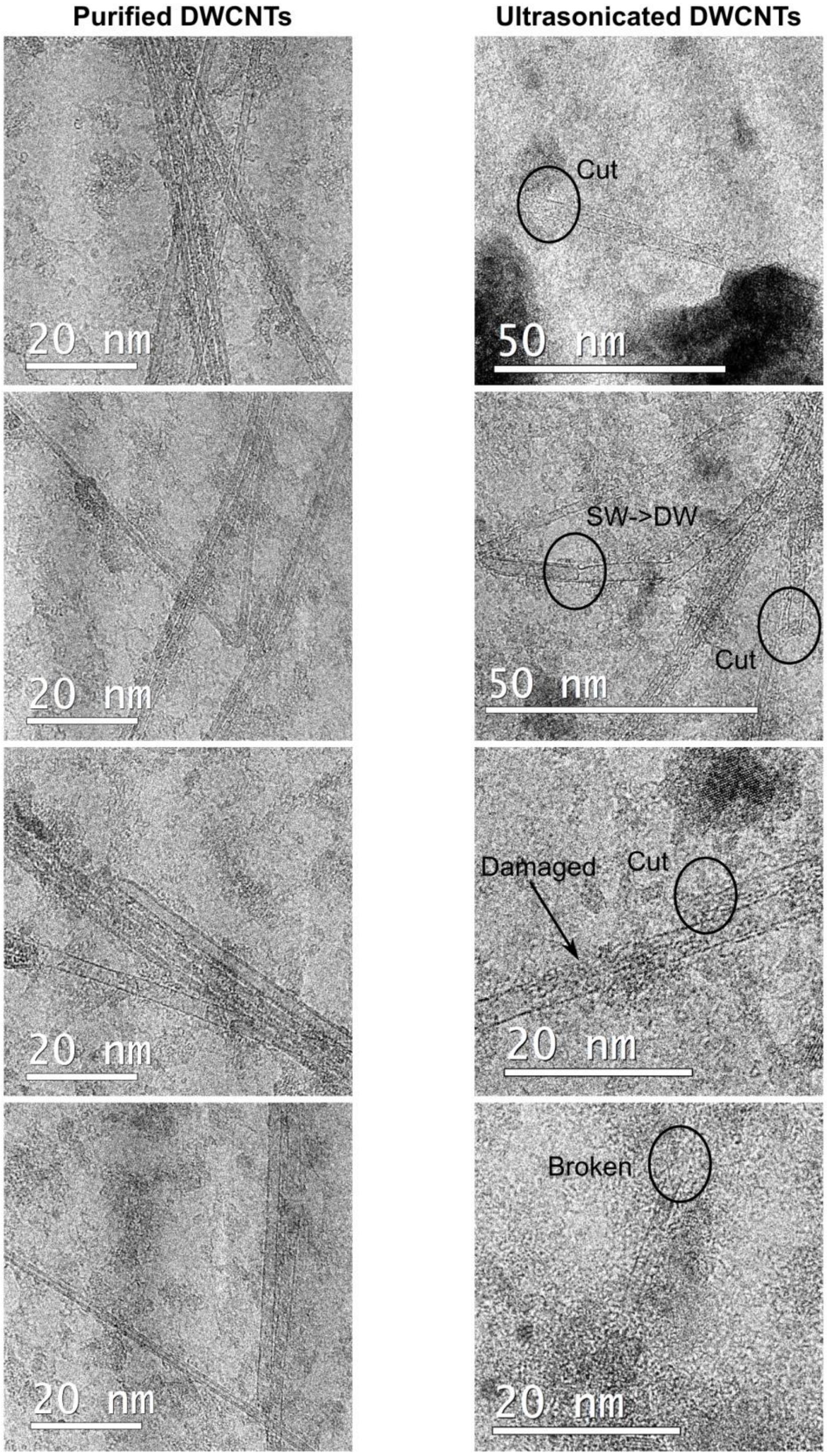

**Figure S19.** Overview of several representative HRTEM images of the purified (left) and ultrasonicated (right) DWCNT sample at high magnification, illustrating how the CNTs are damaged and cut in case of the ultrasonicated DWCNT sample, adding to the already lowered optical response of the large-diameter outer shell SWCNTs compared to that of the inner shell SWCNTs.